\documentclass[10pt,journal,compsoc]{IEEEtran}
\usepackage[ruled,linesnumbered]{algorithm2e}
\usepackage{ulem}
\usepackage{graphicx}
\usepackage{subfig}
\usepackage{amsfonts}
\usepackage[colorlinks,linkcolor=blue]{hyperref}
\usepackage{amssymb}
\usepackage{amsmath}
\usepackage{url}
\usepackage{multirow}
\usepackage{booktabs} 
\usepackage{makecell}
\usepackage{threeparttable}
\usepackage{bm}
\usepackage{color}
\usepackage{mathtools} 
\usepackage[numbers]{natbib}

\newcommand\eqcite[1]{Equation~\eqref{#1}}
\usepackage{ragged2e}
\renewcommand{\justify}{\leftskip=0pt \rightskip=0pt plus 0cm}

\begin{document}

\title{Self-supervised Hypergraph Representation Learning for  Sociological Analysis}

\author{ Xiangguo~Sun,
         Hong Cheng,
         Bo~Liu,
         Jia~Li, 
         Hongyang~Chen,
         Guandong~Xu,
         Hongzhi~Yin
         
 \IEEEcompsocitemizethanks{\IEEEcompsocthanksitem Xiangguo~Sun, and Hong Cheng are with The Chinese University of Hong Kong. \{xgsun,hcheng\}@se.cuhk.edu.hk.  
        
 \IEEEcompsocthanksitem Bo Liu is with Southeast University, China.  bliu@seu.edu.cn
 
  \IEEEcompsocthanksitem Jia Li is with Hong Kong University of Science and Technology (Guangzhou).  jialee@ust.hk
  
  \IEEEcompsocthanksitem Hongyang~Chen is with Zhejiang Lab, China. dr.h.chen@ieee.org

 \IEEEcompsocthanksitem Guandong~Xu is with University of Technology Sydney, Australia. guandong.xu@uts.edu.au

 \IEEEcompsocthanksitem Hongzhi~Yin is with The University of Queensland, Australia. h.yin1@uq.edu.au

 }
}

\markboth{IEEE TRANSACTIONS ON KNOWLEDGE AND DATA ENGINEERING}
{Xiangguo Sun \MakeLowercase{\textit{et al.}}: Self-supervised Hypergraph Representation Learning for  Sociological Analysis}

\IEEEtitleabstractindextext{%
\begin{abstract}
\justify{
Modern sociology has profoundly uncovered many convincing social criteria for behavioral analysis. Unfortunately, many of them are too subjective to be measured and very challenging to be presented in online social networks (OSNs). On the other hand, data mining techniques can better find data patterns but many of them leave behind unnatural understanding to humans. 
In this paper, we propose a fundamental methodology to support the further fusion of data mining techniques and sociological behavioral criteria. First, we propose an effective hypergraph awareness and a fast line graph construction framework. The hypergraph can more profoundly indicate the interactions between individuals and their environments. A line graph treats each social environment as a super node with the underlying influence between different environments. In this way, we go beyond traditional pair-wise relations and explore richer patterns under various sociological criteria; Second, we propose a novel hypergraph-based neural network to learn social influence flowing from users to users, users to environments, environment to users, and environments to environments. The neural network can be learned via a task-free method, making our model very flexible to support various data mining tasks and sociological analysis; Third, we propose both qualitative and quantitive solutions to effectively evaluate the most common sociological criteria like social conformity, social equivalence, environmental evolving and social polarization. Our extensive experiments show that our framework can better support both data mining tasks for online user behaviors and sociological analysis.
}
\end{abstract}
}
\maketitle

\section{Introduction}\label{sec:intro}

\IEEEPARstart{N}{owadays}, online behavior analysis has become one of the most eye-catching interdisciplinary areas in academic and business because of its important role in various web-based applications \cite{li_hyperbolic_2021, wei_beyond_2017, liu_co-detection_2020}. How to fully integrate the most classic achievements from various disciplines is a long-term concern, which can epitomize the whole effort for human-centered computing.

\begin{figure}[t]
    \centering
    \subfloat[social equivalence.]{
     \label{fig:social_equ}
     \includegraphics[width =0.23\textwidth]{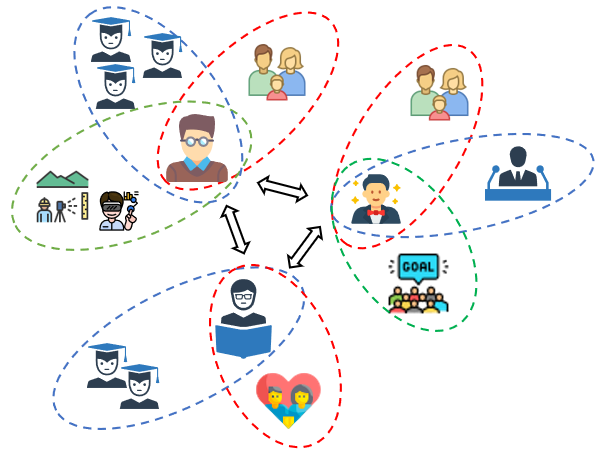}
     }
     \subfloat[social conformity.]{
     \label{fig:social_con}
     \includegraphics[width =0.23\textwidth]{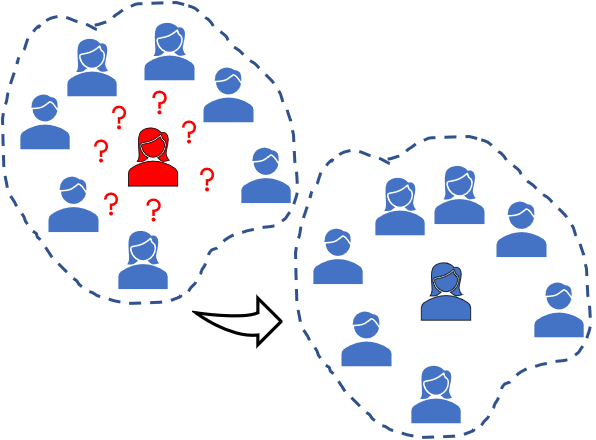}
     }
    \caption{Social Equivalence and Social Conformity}\vspace{-0.2in}
   \label{fig:equ_con}
\end{figure}

The earliest studies on human behavior mostly come from sociologists \cite{myers_twenge_2022} where the most typical theories therein are social equivalence \cite{patrick_doreian_equivalence_1988, hanneman2005introduction, newman_networks_2010} and social conformity \cite{cialdini2004social, cherng_understanding_2022}. As illustrated in Figure \ref{fig:social_equ}, social equivalence measures the sociological distance of a pair of individuals such as the user's social roles, network positions, friendships, etc., and found that two similar persons usually live in a similar social environment. Social conformity states that social influence usually makes people assimilable in their groups and thus causes group members to follow the behavior of the majority. As shown in Figure \ref{fig:social_con}, violating the norm in a social group may yield group pressure \cite{deutsch1955study} to a person, especially when he/she looks for an informative guide to the group.
Social equivalence and social conformity are two skeleton views for learning users' behaviors in the sociology discipline because the former indicates us to distinguish/characterize two individuals via their environments and the latter indicates us to identify different social environments via their members. They are very natural to explain various behaviors online and promising to guide our real-world life. Unfortunately, although being studied for a long time by sociologists, behavior analysis still remains an open problem, especially when it meets online social networks (OSNs) because quantifying these criteria in OSNs is very subjective. Besides, larger data volume from OSNs makes it more difficult to uncover these social criteria than the offline survey.

\begin{figure}[t]
    \centering
    \includegraphics[width =0.38\textwidth]{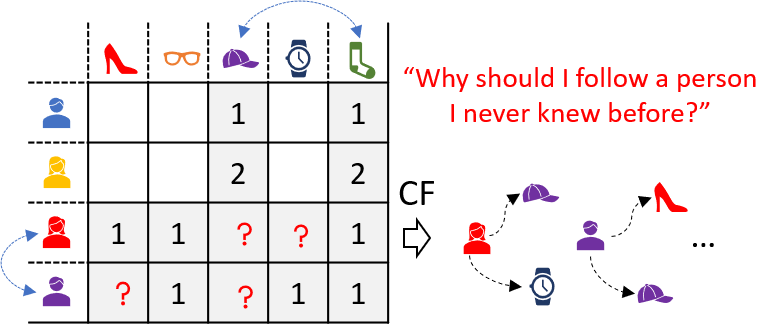}
    \caption{Collaborative Filtering (CF): A widely used data mining technique in online behavior analysis.}\vspace{-0.2in}
    \label{fig:cf}
\end{figure}

The most notable difference between sociological studies and data mining methods for online behavior analysis is that the latter usually characterizes people in various latent variable spaces. Data mining methods claim that users' vectorized representations in latent spaces should also preserve the similarities reflected in the real world. Figure \ref{fig:cf} presents collaborative filtering (CF) \cite{xu_rethinking_2021}, which is one of the most typical data mining paradigms and still dominates data mining area by many frontier researches \cite{yu_self-supervised_2022, yang_hrcf_2022}. It aims at projecting users or items as low-dimensional vectors via Matrix Factorization \cite{guo_deepfm_2017} or more advanced deep neural networks \cite{xie2021graph, chen_neural_2021} on their relation table such as rating, buying, staring, etc. The underlying idea is to find similar users or items in the latent space and then predict a close item for a target user. Indeed it is a very beautiful and clear methodology, but still unnatural for human understanding and less convictive than sociology research. After all, it is far-fetched to force unknown users that have never seen before to believe they are similar in the real world merely because their latent vectors are close, which is far from straightforward compared with various widely accepted sociological theories.

The key problem is: no matter how hard we try, these latent representations only preserve partial similarities reflected in the real world and are far from sufficient because of the limitation of data and models. We can not make the real person fully understand why these latent vectors are meaningful because they are only built on similarity metrics rather than more natural sociological criteria.

Currently, data mining models still suffer from the insufficiency of natural proof for their results although there are lots of work trying to push it forward. Some recent researchers have started to consider sociological concepts on various measurement definitions \cite{tang_confluence_2013, jiang_retweeting_2016}. However, these measurements are so subjective and heavily rely on hand-craft features, which is very hard to be widely applied. 
To this end, other researchers avoid quantifying specific concepts but try to imitate social phenomena such as social influence \cite{qiu2018deepinf, feng_inf2vec_2018}  on pair-wise graphs. Although these models integrate social observations into their models, social influence, from the perspective of sociologists, is far more complicated than pair-wise interactions because it is also widely reflected in various environments. 
Even if we learned individual representations, we still do not understand how these environments perform because there is no corresponding environment uncovered. There are also some works trying to fuse social impact for specific tasks \cite{yin_social_2019, wang2020social}. But applying them to other scenarios might be difficult because they do not provide a task-free framework to capture more general sociological criteria.  
Filling the huge gap between sociology and data mining in online behavior analysis is never easy and needs to solve many tough challenges, including but not limited to:
\begin{itemize} 
    \item How to explore complicated social environments beyond pair-wise relations since most online social networks have not directly indicated them? 
    \item How to design a task-free method to learn social influence flowing that support both data mining and sociological analysis? 
    \item How to evaluate sociological criteria in a data mining model since they are too subjective to be measured?
\end{itemize}

In this paper, we propose a fundamental methodology that can well model online social influence flowing under sociological criteria. Specifically, to address the \textbf{first challenge}, we take both sociological attributes and user connectives with hypergraphs beyond traditional pair-wise relations. A hypergraph allows one edge (a.k.a hyperedge) to connect more than two nodes, making our learned social environments more flexible than classic graphs; To address the  \textbf{second challenge}, we propose a hypergraph-based social influence flowing model to learn individual and environmental interaction patterns under social criteria; To address the \textbf{third challenge}, we propose a package solution to evaluate some widely concerned sociological criteria with both \textbf{qualit}-ative and \textbf{quanti}-tive ways. In summary, our main contributions are as follows:

\begin{itemize}
	\item We propose a hypergraph-based social environment awareness method and a fast line graph algorithm. The hypergraph naturally supports learning the interaction patterns between individuals and their environments. The line graph takes each environment as a new node and learns the interactions between different environments. In this way, our perceived environments are far richer than traditional pair-wise relations and closer to the real world.
	\item We propose a novel hypergraph neural network to learn influence flowing under social criteria. Besides pair-wise relations that can only model the individual impact, our model can further learn social influence flowing from users to environments, environments to users, and environments to environments, making our model more expressive to reflect more profound sociological criteria. 
	\item We propose both qualitative and quantitive methods to evaluate the most common sociological criteria like social conformity, equivalence, environment evolving, social polarization and et al.
	\item We conduct extensive experiments, from which we can find that our framework not only supports better performance in data mining tasks but also effectively preserve rich criteria in sociological analysis.
\end{itemize}

\begin{figure*}[t]
    \centering
    \includegraphics[width =0.8\textwidth]{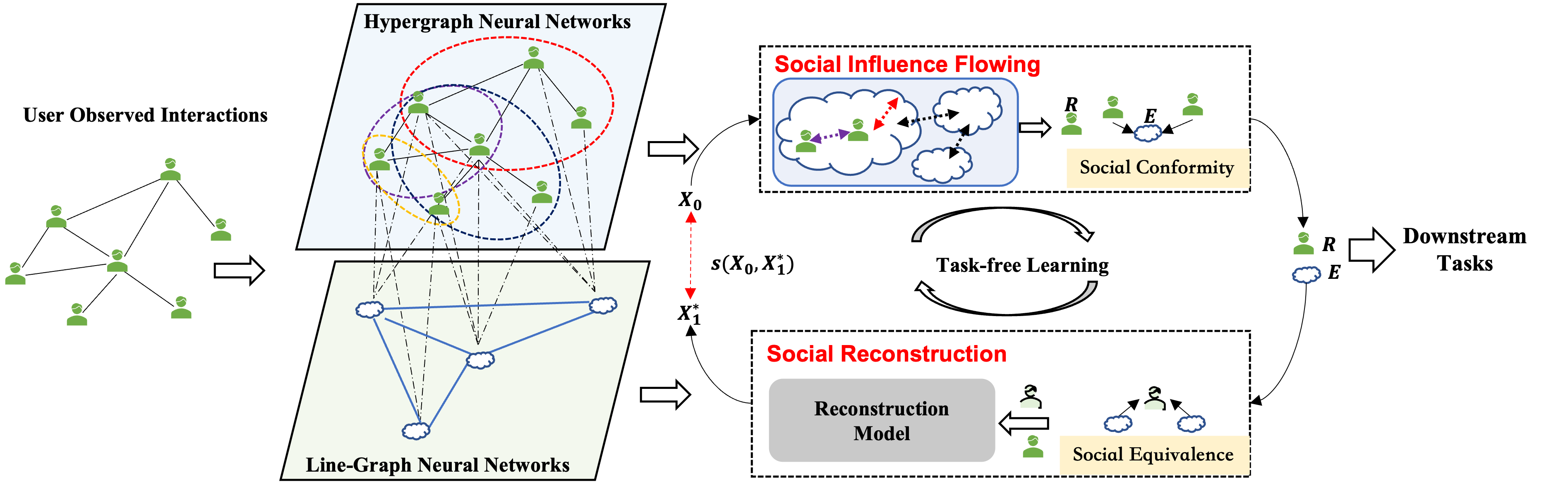}
    \caption{Hypergraph-based Social Influence Flowing under Sociological Criteria.}\vspace{-0.2in}
    \label{fig:framework}
\end{figure*}

\vspace{-0.2in}
\section{Objectives and Definitions}
\newtheorem{definition3}{Definition}
\newtheorem{problem}{Problem}

\begin{definition3}
\textbf{(Hypergraph and its Line Graph)} We denote a hypergraph by $\mathcal{G}=\{\mathcal{V},\mathcal{E}^h,\mathcal{E}^p,\boldsymbol{X},\boldsymbol{W},\boldsymbol{U}\}$ where $\mathcal{V}$ is the node set denoting all the users; $\mathcal{E}^h$ is the collection of hyperedges where each hyperedge can be treated as a set of users this hyperedge contains; $\mathcal{E}^p$ contains pair-wise relations between users; $\boldsymbol{X} \in \mathbb{R}^{|\mathcal{V}|\times d}$ is user attributes matrix; $\boldsymbol{W} \in \mathbb{R}^{|\mathcal{E}^h|\times |\mathcal{E}^h|}$ and $\boldsymbol{U}\in \mathbb{R}^{|\mathcal{V}|\times |\mathcal{V}|}$ are two diagonal matrices where diagonal entries denote the hyperedge weight and user weight, respectively. The line graph of this hypergraph is denoted by $\mathcal{G}^{\circ}=\{\mathcal{V}^{\circ},\mathcal{E}^{\circ}\}$ where each node in $\mathcal{V}^{\circ}$ corresponds to one hyperedge of the hypergraph $|\mathcal{V}^{\circ}|=|\mathcal{E}^h|$; $\mathcal{E}^{\circ}$ is the pair-wise connection of different node in $\mathcal{V}^{\circ}$. Usually, $\mathcal{E}^{\circ}$ exists because hyperedges have some overlapping users or share similar characteristics.
\end{definition3}

In this paper, we use hyperedges to present social environments, and pair-wise links present individual impact. Specifically, we define the social environment as a set of people who have close social patterns and can influence each other, which can be well undertaken by our hyperedges. The hyperedge weights $\boldsymbol{W}$ can be pre-defined as the hyperedge size and the user weights $\boldsymbol{U}$ can be pre-defined as their centrality\footnote{\url{https://wikipedia.org/wiki/Centrality}} in the real world. They can also be manually assigned by the social demand or treated as free parameters learned by downstream tasks.

\textbf{Objectives:} With the above concepts, the objectives of this paper are two-fold: The first one is to design a task-free framework to learn user representation that can reflect social criteria well. Based on this, our second objective is to evaluate whether these representations fused with social criteria perform better in online behavior analysis.

\vspace{-0.1in}
\section{Influence Flowing under Social Criteria }
Our framework is shown in Figure \ref{fig:framework} where the initial online social network is augmented by two layer graphs. The upper graph contains various hyperedges, which contain complicated impact of social environment and users; the bottom graph is the line-graph of the hyperedges, which can present the correlations of different social environments. With the above networks, we propose a hypergraph neural network following sociological influence flowing patterns.

\vspace{-0.1in}
\subsection{Social Environment Awareness via Hypergraphs}

\begin{figure}[h]
\centering
\subfloat[cluster]{
\label{fig:l3}
\includegraphics[width=0.14\textwidth]{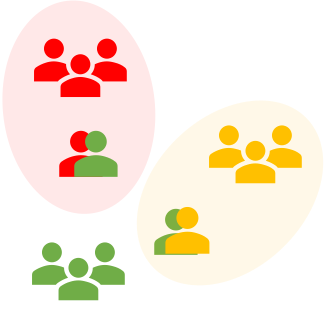}
}
\subfloat[community]{
\label{fig:l3}
\includegraphics[width=0.13\textwidth]{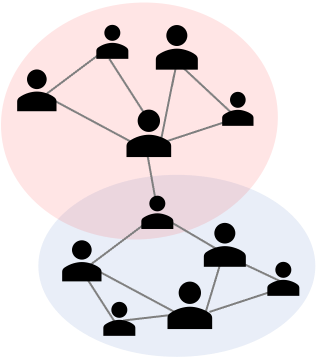}
}
\subfloat[neighbours]{
\label{fig:l3}
\includegraphics[width=0.14\textwidth]{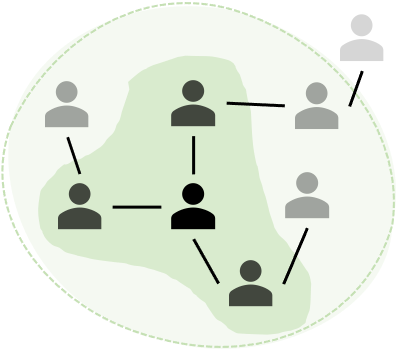}
}
\caption{Commonly Used Social Environments}\vspace{-0.2in}
\label{fig:env}
\end{figure}

Figure \ref{fig:env} presents widely used hyperedge construction methods in existing hypergraph researches. Unfortunately, none of them are sufficient to support more profound sociological criteria. Specifically,  clustering-based methods \cite{zhang2020hypergraph} heavily rely on users' attributes but ignore their real world interactions, making the hyperedges far from flexible. Community-based methods \cite{an2021hypergraph} consider user's network connectivity but in practice not every user should be assigned to a specific community and the hyperedges are less hierarchical; Neighbour-based methods \cite{feng2019hypergraph} take k-hop neighbours around a centered user as a group but this may cause huge hyperedge redundancy. 

Here we propose a new method to reduce the hyperedge redundancy, make them more hierarchical, and let them reflect both social attributes and social interactions. Let $\boldsymbol{F} \in \mathbb{R}_{[0,1]}^{N\times C}$ be user probability distributions on the hyperedges where $N$ is the number of user; $C$ is the number of social environments (a.k.a hyperedges); and each entry value $0\leq \boldsymbol{F}_{ij}\leq 1$. We claim that the hyperedges should both relate to user attributes $\boldsymbol{X} \in \mathbb{R}^{N\times d}$ and their connectivity matrix $\boldsymbol{A} \in \{0,1\}^{N\times N}$. Then $\boldsymbol{F}$ can be estimated by a parameterized ($\theta$) transformation $\overline{\boldsymbol{F}}=f(\boldsymbol{X}|\theta)$ with the following negative log-likelihood target:
\begin{equation}
\small
\begin{aligned}
    \theta^{\star}=\underset{\theta}{\arg \min }\Big\{
    &-\mathbb{E}_{p_{uv}^{+}}\left[\log \left(1-\exp (-f_u(\boldsymbol{X}|\theta)\cdot f_v^{T}(\boldsymbol{X}|\theta))\right)\right]\\
    &+\mathbb{E}_{p_{uv}^{-}}\left[f_u(\boldsymbol{X}|\theta)\cdot f_v^{T}(\boldsymbol{X}|\theta)\right]\Big\}
    \end{aligned}
\end{equation}
where $p^{+}$ and $p^{-}$ denote the probabilities of positive edges and negative edges. Here we can sampled these edges from the connectivity matrix $\boldsymbol{A}$. With the learned hyperedge probability $\boldsymbol{F}$, the social environments can be finally denoted by a hypergraph: $\mathcal{G}=\{\mathcal{V},\mathcal{E}^h,\mathcal{E}^p,\boldsymbol{X},\boldsymbol{W},\boldsymbol{U}\}$.

\vspace{-0.15in}
\subsection{Hypergraph Influence Flowing under Social Criteria }
Unlike most related work that only care about individual impact, more sociological criteria are reflected in both individuals and environment level. To this end, we first generate a line graph of the hypergraph, which can indicate the influence among various environments, then with both hypergraph and its line graph, we propose a hypergraph flowing model to capture richer social criteria.

\vspace{-0.1in}
\subsubsection{Generative Dual Network}
Although our hypergraph contains the potential social patterns of individual to individual and individual to hyperedge, it still lacks of the social patterns of hyperedge to hyperedge, which can be well solved by hypergraph's line graph. Here the line graph takes each hyperedge as a new node and calculates the correlations of different hyperrdges to decide the pair-wise connectivity of the line graph. Let $s_{ij}$ be the similarity score between hyperedge $i$ and $j$. 
Since groups with similar proposition usually cooperate more closely than the others, we can follow the Bernoulli distribution to reconstruct their real world connection as:
\begin{equation*}
\small
    \boldsymbol{A}^{\circ}_{ij}\sim \operatorname{Bernoulli}\left(s_{ij}\right)
\end{equation*}
where $\boldsymbol{A}^{\circ}\in \mathbb{R}^{|\mathcal{E}^h|\times |\mathcal{E}^h|} $ is the estimated adjacency matrix of the line graph. To calculate the similarity score $s_{ij}$ of all pairs of hyperedges, previous work\cite{sun2021multi, yu_self-supervised_2022} usually calculate the Jaccard Similaritiy on any hyperedge pair, or take the transition on hypergraph incidence matrix like $(\boldsymbol{H}^T \boldsymbol{H})_{|\mathcal{E}^h| \times |\mathcal{E}^h|}$. Unfortunately, the first method is low efficient and the second takes up more memories. Even we efficiently get all pairs similarity scores, the total generative space still remains as $|\mathcal{E}^h|\times |\mathcal{E}^h|$, making the generation very slow. 

To solve these problems, we propose a very effective method shown in Algorithm \ref{alg:linegraph} to fast estimate a line graph. Let $\mathcal{V}^{\circ}$ be a set of super nodes where each node corresponds to a hyperedge. Then we can treat the original hypergraph as a bipartite graph $\mathcal{M}=\{\mathcal{V},\mathcal{V}^{\circ},\mathcal{E}^{b}\}$. We start from each super node by selecting a member in this hyperedge and then visit another hyperedge (super node) this member also belong to. We repeat this process until there is no more appropriate nodes selected or the walk length exceeds the max length. Then we can start a new walk and continue the above process. Once the algorithm finished, we can immediately obtain a 
collection of walking paths where each pair of adjacent nodes in a path denotes the concurrence of two hyperedges. The frequency of a hyperedge pair $(i,j)$ on the total collection is an estimation of $s_{ij}$. Therefore, we can directly sampled hyperedge pairs from this collection as the edges of the line graph.

\normalem 
\vspace{-0.1in}
\begin{algorithm}[h]
\SetAlgoNoLine 
\KwIn{Hypergraph $\{\mathcal{V},\mathcal{E}^h\}$;
max length $\ell$;
repeating times $r$;
}
\KwOut{
line graph $\{\mathcal{V}^{\circ}, \mathcal{E}^{\circ}\}$
}
// Initialize a path multiset\\
$\mathcal{C}=\emptyset$\\
//Denote the hypergraph as a bipartitle graph.\\ 
$\mathcal{M}=\{\mathcal{V},\mathcal{V}^{\circ},\mathcal{E}^{b}\}$\\
\For{$i$ from $1$ to $r$}
{
\For{$v_0^{\circ} \in \mathcal{V}^{\circ}$}
{
//random walk on $\mathcal{E}^{b}$ from $v_0^{\circ}$ within $2\ell$ steps.\\
$p=\{v_0^{\circ},v_0,v_1^{\circ},v_1,\cdots,v_0,v_k^{\circ},v_k \}, k\leq \ell$\\
//update the multiset\\
$\mathcal{C}=\mathcal{C}\cup\{(v_0^{\circ},v_1^{\circ}),(v_1^{\circ},v_2^{\circ}),\cdots,(v_{k-1}^{\circ},v_k^{\circ})\}$\\
}
} 
//sample different elements from $\mathcal{C}$\\
$\mathcal{E}^{\circ}\sim P_{(v_i^{\circ},v_j^{\circ})}=\frac{\underset{(v_m^{\circ},v_n^{\circ})\in \mathcal{C}}{\sum}I\left((v_m^{\circ},v_n^{\circ})=(v_i^{\circ},v_j^{\circ})\right)}{|\mathcal{C}|}$\\
\Return{
$\{\mathcal{V}^{\circ}, \mathcal{E}^{\circ}\}$
}
\caption{Fast Line Graph Construction}
\label{alg:linegraph}
\end{algorithm}
\ULforem 

\vspace{-0.2in}
\subsubsection{Hypergraph Neural Network under Social Criteria}
Compared to traditional models that usually consider one-way influence like $group\!\rightarrow \!user$ or $user\!\rightarrow \!user$, we propose to joint the hypergrpah and its line graph to provide more flexible expressiveness on richer social criteria among individual $\Leftrightarrow$ individual, individual $\Leftrightarrow$ environments, and environments $\Leftrightarrow$ environments. To this end, we here propose a novel hypergraph-based neural network to capture social influence flowing.

\textbf{A. Influence Flowing on the Hypergraph.}
Compared with traditional graph models which seek smoothness on pair-wise links, hypergraphs are more feasible to achieve the smoothness on various environmnets, which means we can easily preserve some sociological criteria such as social conformity et al. and perform better with the impact of multiple environments. 

We define a binary hypergraph incidence matrix as $\boldsymbol{H}\in \{0,1\}^{|\mathcal{V}| \times |\mathcal{E}^h|}$ where $\boldsymbol{H}_{uk}=1$ means user $u$ is in the hyperedge $k$ and $\boldsymbol{H}_{uk}=0$ means the user $u$ does not belong to $k$. Let $\boldsymbol{D}^e\in \mathbb{R}^{|\mathcal{E}^h| \times |\mathcal{E}^h|}$ be the hyperedge degree with diagonal entries $\boldsymbol{D}^e(k)=\sum_{u \in \mathcal{V}}\boldsymbol{U}_u \boldsymbol{H}_{uk}$. Similarly, let $\boldsymbol{D}^v(u)=\sum_{k \in \mathcal{E}^h} \boldsymbol{W}_k \boldsymbol{H}_{uk}$ be node degrees; Inspired by social influence theory, social influence flowing from different nodes via multiple environments can be approximated in the following way:
\begin{equation}
\boldsymbol{\Theta}=\left(\boldsymbol{D}^v\right)^{-\frac{1}{2}} \boldsymbol{U} \boldsymbol{H} \boldsymbol{W}\left(\boldsymbol{D}^e\right)^{-1} \boldsymbol{H}^{\mathrm{T}} \boldsymbol{U}\left(\boldsymbol{D}^v\right)^{-\frac{1}{2}}
\end{equation}

Intuitively, $\boldsymbol{H} \boldsymbol{H}^{\mathrm{T}}$ \textbf{\textit{\dotuline{measures the social influence flowing from users to environments and then environments to users}}.} Then $\boldsymbol{\Theta}$ can be treated as a normalized form of $\boldsymbol{H} \boldsymbol{H}^{\mathrm{T}}$ weighted by user and environment impacts. Furthermore, we can extend the above equation to capture farther impact as follows:
\begin{equation}
    \boldsymbol{\Theta}_\Sigma=\sum_{k=1}^K \gamma^{k-1}\boldsymbol{\Theta}^k
\end{equation}
where $\gamma$ is a decay parameter for the observation that far away social influence usually have more feeble impact on users. With the above formulas, user representations on a hypergraph can be affected as follows:
\begin{equation}
    \boldsymbol{X}^{\ell+1}=f\left((\boldsymbol{\Theta}_{\Sigma}-\operatorname{Diag}_{\boldsymbol{\Theta}_{\Sigma}}+\boldsymbol{I}) \cdot \boldsymbol{X}^\ell \cdot \boldsymbol{P}^{\ell}\right)
\end{equation}
where $\operatorname{Diag}_{\boldsymbol{\Theta}_{\Sigma}}$ is the diagonal matrix of $\boldsymbol{\Theta}_{\Sigma}$; $\boldsymbol{I} \in \{0,1\}^{|\mathcal{V}|\times |\mathcal{V}|}$ is the identity matrix; $\boldsymbol{P}^{\ell} \in \mathbb{R}^{|\mathcal{V}|\times d}$ is the feature transformation matrix. $f(\cdot)$ is an activate function such as ``Relu'', ``Sigmoid'' et al; $\ell=1,\cdots, L$ denotes the neural network layer; The last layer output can be denoted by $\boldsymbol{R}^h=\boldsymbol{X}^{L+1}$.

\textbf{B. Influence Flowing on Pair-wise Relations.}
Besides the interactions between users and their environment, our hypergraph notation also includes the pair-wise links $\mathcal{E}^p$, which can support us build individual level impact via simple graph neural network \cite{kipf2017semi} like:
\begin{equation}\label{equ:gcn}
(\boldsymbol{X}^p)^{\ell+1}=f\left((\boldsymbol{D}^p)^{-\frac{1}{2}} \boldsymbol{A}^p (\boldsymbol{D}^p)^{-\frac{1}{2}} (\boldsymbol{X}^p)^{\ell} (\boldsymbol{P}^p)^{\ell}\right)
\end{equation}
where $\boldsymbol{A}^p \in \mathbb{R}^{|\mathcal{V}|\times |\mathcal{V}|}$ is a pair-wise adjacency matrix built on $\mathcal{E}^p$ and self-connections; $\boldsymbol{D}^p_{i i}=\sum_j \boldsymbol{A}^p_{i j}$; $\boldsymbol{P}^p$ is the weight matrix. The last layer output can be denoted by $\boldsymbol{R}^p=(\boldsymbol{X}^p)^{L+1}$. Intuitively, \eqcite{equ:gcn} \textbf{\textit{\dotuline{measures the social impact in individual level}.}} With user representations both from hypergraph and pair-wise relations, we can combine them together:
\begin{equation}
    \boldsymbol{R}^{encode}=\boldsymbol{R}^h\oplus\boldsymbol{R}^p
\end{equation}
where $\oplus$ is the concatenation operation.

\textbf{C. Influence Flowing on the Line Graph.}
Besides the above social influence flowing, social criteria are also widely reflected in mutual influence between different environments, which can be well addressed in hypergraph's line graph $\mathcal{G}^{\circ}=\{\mathcal{V}^{\circ},\mathcal{E}^{\circ}\}$ where each node in $\mathcal{V}^{\circ}$ corresponds to an environment entity in the hypergraph and $\mathcal{E}^{\circ}$ are pair-wise interactions for  these environments. 

Inspired by social conformity theory, we can represent the environment by taking the summation of all its members like $\boldsymbol{X}^{\circ}=\boldsymbol{H}^{T}\boldsymbol{R}^{encode}$ where $\boldsymbol{X}^{\circ}\in \mathbb{R}^{|\mathcal{V}^{\circ}|\times d}$ is the hyperedge representation matrix. Then we can \textbf{\textit{\dotuline{measure the social influence flowing from an environment to another environment}}} on the line graph by a similar graph neural network to \eqcite{equ:gcn}. Let the updated hyperedge representations be $(\boldsymbol{X}^{\circ})^{L+1}$. Inspired by social equivalence theory, we can use the environments to denote its members as: $\boldsymbol{R}^{\star}=\boldsymbol{H}(\boldsymbol{X}^{\circ})^{L+1}$ where $\boldsymbol{R}^{\star} \in \mathbb{R}^{|\mathcal{V}|\times d}$ is the reconstructed user representations from their environments.

\vspace{-0.15in}
\subsection{Task-free Learning by Dual and Contrastive Ideas}


Contrastive learning ~\cite{yu_self-supervised_2022} on simple graphs usually generates different graph views (different topological structures) and then use the same encoder to get their own representations. It tries to make the same entity over different views as close as possible and keeps nonaligned entities away from each other. Unfortunately, although our framework might generate several ``views'' (hyperedges, pair-wise relations, and line graph edges), they can not be modeled by the same encoder, and nodes in different views are not always aligned (for example, nodes in line-graph are not aligned with any nodes in other graphs). Besides, there are also rich influence between different views neglected, leaving classic contrastive methods not applicable here. Luckily, we can borrow recent idea of dual learning~\cite{he2016dual} in machine translation that uses two dual translators (like \begin{scriptsize}${ Chinese\!\rightarrow \!English}$\end{scriptsize} and \begin{scriptsize}${ English\!\rightarrow \!Chinese}$\end{scriptsize}) to compare the initial representations and the reconstructed ones. In this paper, we can also obtain users' initial features and their reconstructed representations from the hypergraph and its line graph. However, dual learning only cares about the same entity but ignores the other entity pairs that also impact the final learning results.

To this end, we first obtain the reconstructed user representation by projecting previous $\boldsymbol{R}^{\star},\boldsymbol{R}^h, \boldsymbol{R}^p$ via a Multilayer Perceptron (MLP) like: $\widehat{\boldsymbol{X}}=\operatorname{MLP}(\boldsymbol{R}^{\star}\oplus\boldsymbol{R}^h\oplus\boldsymbol{R}^p)$. Then we compare users' initial representation $\boldsymbol{X}$ and the reconstructed matrix $\widehat{\boldsymbol{X}}$ with both aligned entities a.k.a positive pairs $\mathcal{P}=\{(u,u)|u\in \mathcal{V}\}$ and nonaligned entity pairs a.k.a negative pairs $\mathcal{N}=\{(u,v)|u\in \mathcal{V},v\in \mathcal{V},(u,v)\notin \mathcal{E}^p\}$. Our task-free training loss with both feature and structure reconstruction can be defined as follows: 
\begin{equation}\label{equ:loss}
\small
     \mathcal{L}\!=\!\!\!\!\!\!\underset{(u,u)\in \mathcal{P}}{\sum}\!\!\!\left[\|\boldsymbol{X}_u\!-\!\widehat{\boldsymbol{X}}_u\|^2_2\!-\!m_p\right]_{+}
     \!\!\!\!+\!\!\!\!\underset{(u,v)\in \mathcal{N}}{\sum}\!\!\!\!\left[m_n\!\!-\!\|\boldsymbol{X}_u\!\!-\!\widehat{\boldsymbol{X}}_v\|^2_2\right]_{+}
\end{equation}
where $m_p$ presents the margin over which positive pairs will contribute to the loss; $m_n$ is the negative margin under which negative pairs will contribute to the loss. After the learning process finished, we can use the updated model to generate $\boldsymbol{R}^{\star},\boldsymbol{R}^h, \boldsymbol{R}^p$ to support downstream tasks, which can be vividly named as ``pluggable'' way. We can also combine \eqcite{equ:loss} with task specific loss and then directly conduct downstream tasks in an ``unpluggable'' way.

\vspace{-0.1in}
\subsection{Evaluating Sociological Criteria}
There is a long-term concern on how to reveal the sociological criteria in a data mining model because they are usually too subjective to be measured. Here, we wish to push forward by a package solution to some widely concerned sociological criteria:

\textbf{Social Conformity. }
With the representations of users and hyperedges learned by the model, 
social conformity can be easily visualized by the users who have higher similarities to the hyperedge (section \ref{task:equ_con}). Besides, it can be also quantitatively measured as follows:
\begin{equation}
\small 
    co=\mathbb{E}_{e\sim P(\mathcal{E}^h)}\left[\sum_{u\in e}\frac{\operatorname{I}(s_{ue}>\rho)}{|e|}\right]
\end{equation}
where $co$ is the expected ratio of significant users to each hyperedge; $\operatorname{I}(\cdot)$ is the condition indication; $s_{ue}$ is the similarity of user $u$ and hyperedge $e$; $\rho$ is the significance threshold. $P(\mathcal{E}^h)$ is the probability distribution on hyperedges.

\textbf{Social Equivalence.} A visualization way is to draw the hyperedges and compare whether close users in the real world also share similar hyperedges (section \ref{task:equ_con}), which is also equivalent to a quantified measurement as follows:
\begin{equation}
    \footnotesize
    eq\!=\!\frac{\!\!\!\!\!\!
    \underset{(u,v) \sim P(\mathcal{E}^p)}{\mathbb{E}}
    \left[
        \underset{e \in \mathcal{E}^h}{\sum}
        \!\!\!\operatorname{I}(u\!\in\! e\! \land\! v\! \in\! e)/\!\operatorname{I}(u\!\in \!e \!\lor\! v \!\in \!e)
    \right]
    }{\!\!\!\!\!\!
    \underset{(u^{'}\!\!,v^{'}) \sim P(\!\mathcal{V}\times \mathcal{V} \setminus \mathcal{E}^p)}{\mathbb{E}}
   \left[
        \underset{e \in \mathcal{E}^h}{\sum}
        \!\!\!\operatorname{I}(u^{'}\!\!\in \!e \!\land \!v^{'}\!\! \in e)/\!\operatorname{I}(u^{'}\!\!\in\! e \!\lor \!v^{'} \!\!\in \!e)
    \right]
    }
\end{equation}
where the numerator is the expected Jaccard index of environments for the positive user pairs; the denominator is the same thing for the negative pairs. Intuitively, $eq>1$ means the social equivalence is significantly observed.

\textbf{Social Environment Evolving.}
We calculate the ratio of significant users to the group size and then we can directly see the ratio changing during different phases (section \ref{task:evol}).  
 
\textbf{Social Polarization.}
As a natural consequence of social influence, social polarization \cite{myers_twenge_2022} is the phenomenon that users may enhance their belief if their neighbours take the same attitudes. Here we measure a user's belief from the perspective of \textit{information theory}\footnote{\url{https://wikipedia.org/wiki/Information_theory}}.
For each node $u$ in a hyperedge $i$, we use the dot product of $u$ and $i$ to present the confidence of this user believing his opinion, say $p_{ui}$. Then the \textit{uncertainty} is defined as $h_{ui}=-\log p_{ui}$ and the lower value indicates more radical opinion this user holds. Then the Social Polarization within a group can be easily visualized by the uncertainty distributions of group members (section \ref{task:pola}). Furthermore, the overall polarization degree in group $i$ can be also measured by the expectation of all group members a.k.a \textit{group entropy}: 
\begin{equation}
    po(i)=-\mathbb{E}_{u \sim p_{ui}}\left[ \log p_{ui}\right]
\end{equation}

\section{experiment}

\subsection{Datasets}
We evaluate our framework on five real-world datasets. Detailed statistics are presented in Table \ref{tab:data}. MovieLens\footnote{available at \url{https://movielens.org}} is a movie rating network where nodes include "movie" and "user". Each observed pair-wise link starts from a user and points to a movie. The link value presents the user's rating score (from $0$ to $5$) to a target movie; Facebook \cite{yang2020scaling} is a social network where nodes denote user accounts and pair-wise edges are following relations; FakeNews \cite{dou2021user} dataset contains 314 retweeting networks where 157 of them are fake news. Nodes in each graph are users and pair-wise links denote the retweeting relations; PersonalityCafe\footnote{available at \url{https://www.personalitycafe.com}} is an online forum where most of the users have their personality types which are denoted by Myers–Briggs Type Indicator (MBTI)\footnote{\url{https://wikipedia.org/wiki/Myers-Briggs_Type_Indicator}}. The dataset contains 42K users as nodes and the pair-wise links denote some users quoted the others in the forum; KaraClub \cite{zachary1977information} is a classic small-scaled network with 34 nodes connected by 156 edges. We use it to visualize our hyperedge structure compared with other methods.

\vspace{-0.1in}
\subsection{Baselines}
We evaluate our framework with the following baselines: metapath2vec \cite{dong2017metapath2vec} projects network entities to multiple latent spaces with independent random walks along different meta paths; Confluence \cite{tang_confluence_2013} first defines social conformity as three kinds of calculations from user's historical actions and then uses the relation network to learn the weights of different kinds of conformity metrics; OCCFRP \cite{jiang_retweeting_2016} quantifies social influence and introduce it with collaborative filtering method to predict users' retweeting actions; NCF \cite{he_neural_2017} enhances collaborative filtering model with deep neural network where each user or item is represented by a neural encoder and the learning process integrate both Matrix Factorization and the multi-layer perceptron; DeepInf \cite{qiu2018deepinf}models social influence within local sub-network uses graph convolution network to encode instances from the sampled neighbors; SIDM \cite{wang2020social} splits social interaction network as influence structure and influence dynamics and then model each of them via graph neural networks.

\begin{table}[h]
\centering
\caption{Dataset Statistics}\vspace{-0.1in}
\label{tab:data}
\resizebox{0.4\textwidth}{!}{%
\begin{tabular}{@{}lllll@{}}
\toprule
                & \makecell[l]{Node \\\#Number}       
                & \makecell[l]{Pair-wise Relation \\\#Number   }         
                \\ \midrule
MovieLens       & \makecell[l]{movie; user\\\#9,742; \#610}  
                & \makecell[l]{(user,rates, movie) \\\#70,586}   
                \\\midrule
Facebook        & \makecell[l]{user \\\# 4,039}       
                & \makecell[l]{(user, follows, user) \\\#88,234} 
                \\\midrule
FakeNews        & \makecell[l]{user \\\#41,054} 
                & \makecell[l]{(user retweet user) \\\#40,740}   
                \\\midrule
PersonalityCafe & \makecell[l]{user \\\# 42,360}    
                & \makecell[l]{(user, quote, user)\\ \#889,706}    
                \\\midrule
KarateClub      & \makecell[l]{user \\\#34}      
                & \makecell[l]{(user, follow, user) \\\#156 }             
\\ \bottomrule
\end{tabular}
}
\end{table}
\vspace{-0.2in}

\subsection{Metrics and Experimental Settings}

\textbf{Tasks and Metrics:} 
For the rating problem, we aim at predicting users' rating scores for specific items (such as movies). We treat this problem as value regression and thus the adopted metrics include MAE (mean absolute error), MSE (mean squared error), R2 (coefficient of determination), RMSE (root mean square error), and Max Error. For the following relation awareness, we treat it as the link prediction and evaluate on AUC (area under the ROC curve), AP (average precision, a.k.a area under the PR curve), MRR (mean reciprocal rank), Hits@1, and Hits@5 (hit ratio at top 1 and top 5). For the fake news identification, each tweet has its own retweeting graph and we treat it as a binary graph classification. Our metrics include AUC, AP, F1, Precision, and Recall. For personality detection, we treat it as a multi-label classification on graph nodes and evaluate the performance by Accuracy, Macro/Micro F1, Macro/Micro Precision, and Macro/Micro Recall.

\begin{table*}[t]
\centering
\caption{Personality Detection ($30\%$ labeled)}\vspace{-0.1in}
\label{tab:personality}
\resizebox{1.8\columnwidth}{!}{
\begin{threeparttable}
\begin{tabular}{@{}lccccccc@{}}
\toprule
                    & Accuracy & Macro F1 &Macro Precision  & Macro Recall & Micro F1 & Micro Precision & Micro Recall \\ \midrule
metapath2vec        & 0.1856   & 0.7450   & 0.7155          & 0.8315       & 0.7645   & 0.7106          & 0.8272       \\
Confluence          & 0.2201   & 0.7567   & 0.7362          & 0.8003       & 0.7573   & 0.7435          & 0.8066       \\
OCCFRP              & 0.2366   & 0.7786   & 0.7603          & 0.8102       & 0.8025   & 0.7701          & 0.8238       \\
NCF                 & 0.2515   & 0.7805   & 0.7764          & 0.8035       & 0.7897   & 0.7805          & 0.8105       \\
DeepInf             & \underline{\textbf{0.2937}}   & 0.7703   & \underline{\textbf{0.7903}}          & 0.7685       & 0.7854   & \underline{\textbf{0.8006}}          & 0.7764       \\
SIDM                & 0.2891   & 0.7819   & 0.7607          & 0.8127       & 0.7974   & 0.7706          & 0.8260       \\ \midrule
Our (pluggable)     & 0.2593   & \underline{\textbf{0.8169}}   & 0.7141          & \underline{\textbf{0.9902}}       & \underline{\textbf{0.8244}}   & 0.7197          & \underline{\textbf{0.9920}}       \\
Our (unpluggable) & \underline{\textbf{0.4555}}   & \underline{\textbf{0.8337}}   & \underline{\textbf{0.8064}}          & \underline{\textbf{0.8695}}       & \underline{\textbf{0.8464}}   & \underline{\textbf{0.8171}}          & \underline{\textbf{0.8837}}       \\ \midrule
(bottom line)       & (0.0625)   & (0.5751)   & (0.6885)          & (0.4999)       & (0.5792)   & (0.6885)          & (0.4999)       \\ \bottomrule
\end{tabular}
\begin{tablenotes}
        \footnotesize
        \item \underline{\textbf{value}}: top 2 results.
      \end{tablenotes}
\end{threeparttable}
}\vspace{-0.2in}
\end{table*}

\noindent \textbf{Data Preparing } 
For the Movielens dataset, we remove 20\% of the total observed rating edges as a testing set, split 10\% edges as validation, and the rest as training edges. Each edge contains a rated score from $0$ to $5$ which will be scaled within $[0,1]$ for training and evaluation. For the Facebook dataset, we split training, validation, and testing edges with the same proportion in Movielens. In the meanwhile, we also sample $100$ negative edges for each positive edge so that we can evaluate the performance on link prediction via AUC, AP, MRR, and Hits@K. For the FakeNews dataset, each independent graph denotes a retweeting network and we treat the task as graph classification. We take the default split by the dataset ($62:31:221$ for training, validating, and testing); For the PersonalityCafe dataset, there are 42, 360 users in total but only 22, 817 of them have MBTI personality labels. We split these 22, 817 users as training, validation, and testing nodes with the proportion of $10\text{K}:2.8\text{K}:10\text{K}$. Users with observed personality labels in the training process only take up 30\% of the total.

\noindent \textbf{Parameter Settings}
For most tasks, we train our model with 200 epochs, Adam optimizer, 0.01 learning rate, and 100 dimension initial entity features obtained from metapath2vec. We take two layers for the classical graph and hypergraph neural networks in the encoding process and one layer of graph neural networks in the decoding process. We design this asymmetric structure so that we can learn stronger representations and reduce training complexity. More detailed settings are presented in our open project\footnote{\url{https://tinyurl.com/tkde2023}}.

\vspace{-0.1in}
\subsection{Behavior Analysis based on our Model}
\label{subsec:behavior}

\subsubsection{Effectiveness on Personality Detection}
\label{task:personality}
In the field of sociological psychology, users' personality traits are the most crucial factors to analyze their behaviors. One of the most acceptable psychology models to denote one's personality traits is the Myers-Briggs Type Indicator, which evaluates users over four categories: introversion/extraversion, sensing/intuition, thinking/feeling, judging/perceiving. Each person has one preferred quality from each category, making the target problem equal to multi-label classification.

We repeat our experiment $10$ times and report the average results of our model and the other baselines in Table \ref{tab:personality}, from which we can observe that our method achieves the best results in most metrics. Here the ``pluggable'' method means we only train our model in a self-supervised way and then load the learned representations to train a downstream multi-label classifier. The ``unpluggable'' method means we join the representation learning and multi-label classifier together and then directly output users' personality traits. Compared with the unpluggable method, our pluggable method also outperforms the rest baselines in most metrics, which can further indicate the effectiveness of our learning framework. Note that SIDM, DeepInf, OCCFRP, and Confluence all try to integrate some social influence observations into their model but SIDM and DeepInf avoid manually defined measurements on these subjective concepts and achieve better performance than the rest. However, social influence is only a narrow aspect of sociology research and there are more profound sociological routines that can further support personality detection in our framework, which are also observable in the following experiments.

\begin{figure*}[t]
\centering
\subfloat[Initial Feature]{
\label{fig:emb:a}
\includegraphics[width=0.25\textwidth]{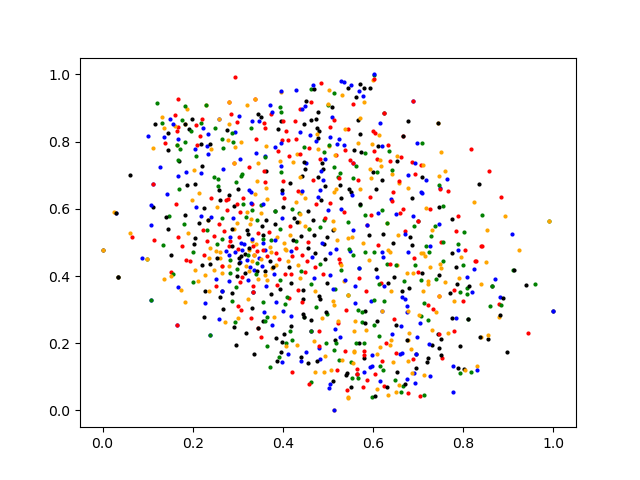}
}
\subfloat[AutoEncoder]{
\label{fig:emb:b}
\includegraphics[width=0.25\textwidth]{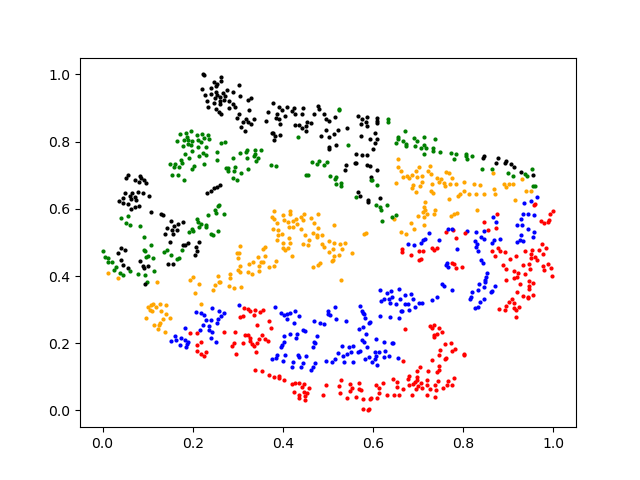}
}
\subfloat[GCN]{
\label{fig:emb:c}
\includegraphics[width=0.25\textwidth]{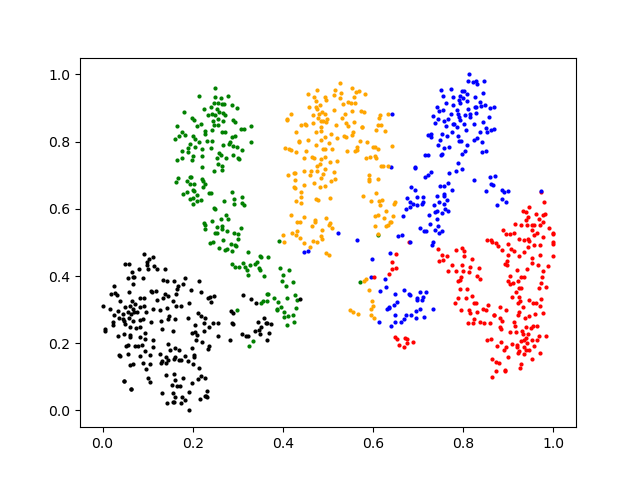}
}
\subfloat[Our Model]{
\label{fig:emb:d}
\includegraphics[width=0.25\textwidth]{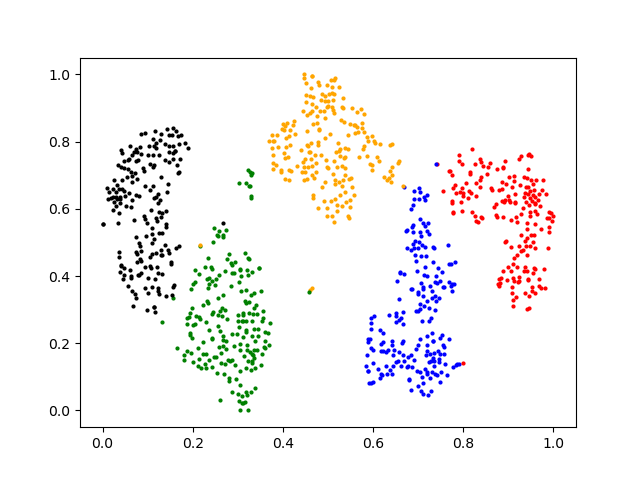}
}
\caption{Representation Distributions w.r.t Learning Models}\vspace{-0.2in}
\label{fig:emb}
\end{figure*}

\subsubsection{Effectiveness on Rating Estimation}
\label{task:rating}
Rating items is one of the most common cases in many data mining applications where users can score the item to present their preferences. Here we treat this task as score value regression and report the average results in Table \ref{tab:rate}. Here the ``pluggable'' method means we only train our framework by our proposed dual contrastive loss and then directly output the predicted score via dot operation on user and item representations. ``unpluggable'' method means we train our framework by the joint of dual contrastive loss and a task-specific loss such as the ``L1Loss'' (a.k.a mean absolute error).

\begin{table}[h]
\centering
\caption{Rating Prediction}\vspace{-0.1in}
\label{tab:rate}
\resizebox{\columnwidth}{!}{
\begin{threeparttable}
\begin{tabular}{@{}lccccc@{}}
\toprule
                    & MAE\tnote{$\downarrow$}    & MSE\tnote{$\downarrow$}    & R2\tnote{$\uparrow$}     & RMSE\tnote{$\downarrow$}   & Max Error\tnote{$\downarrow$} \\ \midrule
metapath2vec        & 0.2476 & 0.0920 & 0.0889 & 0.3033 & 0.9986    \\
Confluence          & 0.3079 & 0.1306 & 0.0065 & 0.3614 & 0.9988    \\
OCCFRP              & 0.2866 & 0.1049 & 0.0926 & 0.3239 & 0.9255    \\
NCF                 & 0.2376 & 0.0901 & 0.1106 & 0.3002 & 0.8972    \\
DeepInf             & 0.2235 & 0.0860 & 0.1303 & 0.2933 & \underline{\textbf{0.8865}}    \\
SIDM                & 0.2011 & 0.0794 & 0.1326 & 0.2818 & 0.8990    \\ \midrule
Our (pluggable)     & \underline{\textbf{0.1457}} & \underline{\textbf{0.0371}} & \underline{\textbf{0.2392}} & \underline{\textbf{0.1925}} & \underline{\textbf{0.8681}}    \\
Our (unpluggable) & \underline{\textbf{0.1397}} & \underline{\textbf{0.0347}} & \underline{\textbf{0.2870}} & \underline{\textbf{0.1863}} & 0.8920    \\ \bottomrule
\end{tabular}
\begin{tablenotes}
        \footnotesize
        \item $\uparrow$: higher better; $\downarrow$: lower better; \underline{\textbf{value}}: top 2 results.
      \end{tablenotes}
\end{threeparttable}
}\vspace{-0.1in}
\end{table}

As shown in Table \ref{tab:rate}, our model still keeps better performance on all the metrics. In particular, our unpluggable method achieves the best results in most cases. Although it has larger Max Error than DeepInf, it is still very competitive. Besides our methods, SIDM and DeepInf also achieve generally better performance than the rest baselines, which may further suggest the advantages of integrating the data mining model with profound sociological observations.

\subsubsection{Effectiveness on Following Relation Awareness}
\label{task:following}
For the task of following relation awareness, we first sample $100$ negative following edges for each positive edge and then use our learned node embedding to discriminate the positive ones from the blended testing edge set. We treat this task as unbalanced binary classification and report the average results in Table \ref{tab:follow} where the ``pluggable'' way of our method means we train our model in a self-supervised way and then load the learned representations to train a binary edge classifier with the cross entropy loss. Since the positive and negative edges are very unbalanced, we take the gradient harmonizing mechanism (GHM) \cite{li2019gradient} to regular the task loss. The ``unpluggalbe'' way of our method means we bland our model embedding with a binary classifier and then jointly train the total pipeline with our proposed dual contrastive loss and the classification loss.

\begin{table}[h]
\centering
\caption{Following Relation Awarenesss}\vspace{-0.1in}
\label{tab:follow}
\resizebox{\columnwidth}{!}{%
\begin{threeparttable}
\begin{tabular}{@{}lccccc@{}}
\toprule
                    & AUC    & AP     & MRR    & Hits@1 & Hits@5 \\ \midrule
metapath2vec        & \underline{\textbf{0.9322}} & 0.2122 & 0.4869 & 0.2385 & 0.7588 \\
Confluence          & 0.8902 & 0.2164 & 0.4991 & 0.2897 & 0.7536 \\
OCCFRP              & 0.8864 & 0.2201 & 0.5115 & 0.3006 & 0.7644 \\
NCF                 & 0.9037 & 0.2334 & 0.5564 & \underline{\textbf{0.3537}} & 0.7967 \\
DeepInf             & 0.9106 & 0.2606 & \underline{\textbf{0.5632}} & 0.3162 & 0.8036 \\
SIDM                & 0.9237 & 0.2564 & 0.5537 & 0.3384 & \underline{\textbf{0.8364}} \\ \midrule
Our (pluggable)     & \underline{\textbf{0.9707}} & \underline{\textbf{0.2678}} & 0.5576 & 0.3508 & \underline{\textbf{0.8589}} \\
Our (unpluggable) & 0.9297 & \underline{\textbf{0.4158}} & \underline{\textbf{0.6198}} & \underline{\textbf{0.4896}} & 0.7805 \\ \bottomrule
\end{tabular}
\begin{tablenotes}
        \footnotesize
        \item \underline{\textbf{value}}: top 2 results.
      \end{tablenotes}
\end{threeparttable}
}\vspace{-0.1in}
\end{table}

Similar to previous tasks, our model achieved very competitive performance. Specifically, the ``pluggable'' way of our method comes to the top in AUC and Hits@5. the ``unpluggable'' way of our method performs the best in AP, MRR, and Hits@1. We also observe that in some metrics such as AUC, MRR, and Hits@1, the rest baselines can also achieve very competitive performance (usually second the best), but we still hold a strong position and keep very close to them. This task further demonstrates that our model is very reliable even in unbalanced data distribution.

\subsubsection{Effectiveness Analysis on Fake News Identification}
\label{task:fakenews}
In online social networks, fake news is usually spread via users' retweeting action, making the identification more like a graph classification task where each graph is a retweeting network and nodes in the graph are users. Here we take the default setting of the FakeNews dataset where each node in the graph has a 310-dimensional feature as input. The initial feature contains 10 Twitter user profile attributes and a 300-dimensional representation of the user's historical tweets encoded by the word2vec model.

\begin{table}[h]
\centering
\caption{Fake News Identification ($30\%$ labeled)}\vspace{-0.1in}
\label{tab:fake}
\resizebox{\columnwidth}{!}{
\begin{threeparttable}
\begin{tabular}{@{}lccccc@{}}
\toprule
                    & AUC    & AP     & F1     & Precision    & Recall    \\ \midrule
metapath2vec        & 0.8873 & 0.8387 & 0.7904 & 0.8582 & 0.7325 \\
Confluence          & 0.8525 & 0.8237 & 0.8149 & 0.8365 & 0.7944 \\
OCCFRP              & 0.8936 & 0.8364 & 0.8857 & \underline{\textbf{0.8972}} & 0.8744 \\
NCF                 & 0.9003 & 0.8537 & \underline{\textbf{0.9053}} & \underline{\textbf{0.9106}} & 0.9001 \\
DeepInf             & 0.9106 & 0.8865 & 0.8751 & 0.8573 & 0.8936 \\
SIDM                & \underline{\textbf{0.9264}} & \underline{\textbf{0.9036}} & 0.9045 & 0.8864 & \underline{\textbf{0.9233}} \\ \midrule
Our (pluggable)     & 0.9170 & 0.8911 & 0.8770 & 0.8688 & 0.8854 \\
Our (unpluggable) & \underline{\textbf{0.9507}} & \underline{\textbf{0.9525}} & \underline{\textbf{0.9125}} & 0.8957 & \underline{\textbf{0.9299}} \\ \bottomrule
\end{tabular}
\begin{tablenotes}
        \footnotesize
        \item \underline{\textbf{value}}: top 2 results.
      \end{tablenotes}
\end{threeparttable}
}\vspace{-0.2in}
\end{table}

As shown in Table \ref{tab:fake}, our models keep the best performance in AUC, AP, F1, and Recall. In particular, the ``pluggable'' way of our method means we first obtain learned node-level embeddings and then get graph-level representation by graph pooling. Afterward, we feed these graph representations to a graph classifier to predict whether it belongs to fake news or not.  The ``unpluggable'' way of our method means we concatenate the graph classifier directly to our node-level learning and jointly train the total workflow via the combination of both self-supervised loss and cross-entropy loss. Although the ``pluggable'' way lost the top 2 positions, it still keeps a very close performance to the others. Since it can learn node embeddings independent from the specific task, we would directly leverage them for other similar tasks without training from scratch.

\vspace{-0.1in}
\subsection{Visualization of Learned Embeddings}
\label{task:vis}

In this section, we select five groups in the PersonalityCafe dataset and compare the discrimination capability of our model with other classic embedding routines. Note that for fair competition, these five groups are not found by our model but the original nature groups in the forum. We draw the representation distributions by t-SNE and present the results in Figure \ref{fig:emb}. The initial features (Figure \ref{fig:emb:a}) are encoded by the metapath2vec model on the pair-wise links with meta path ``user-quote-user'', and then we feed them to the rest routines such as AutoEncoder (Figure \ref{fig:emb:b}), GCN (Figure \ref{fig:emb:c}), and our model (Figure \ref{fig:emb:d}). From the comparison, we can find that our model learns higher-quality representations for the users and the resolution is better than the rest routines.

\begin{figure*}[h]
\centering
\subfloat[cluster]{
\label{fig:structure:a}
\includegraphics[width=0.24\textwidth]{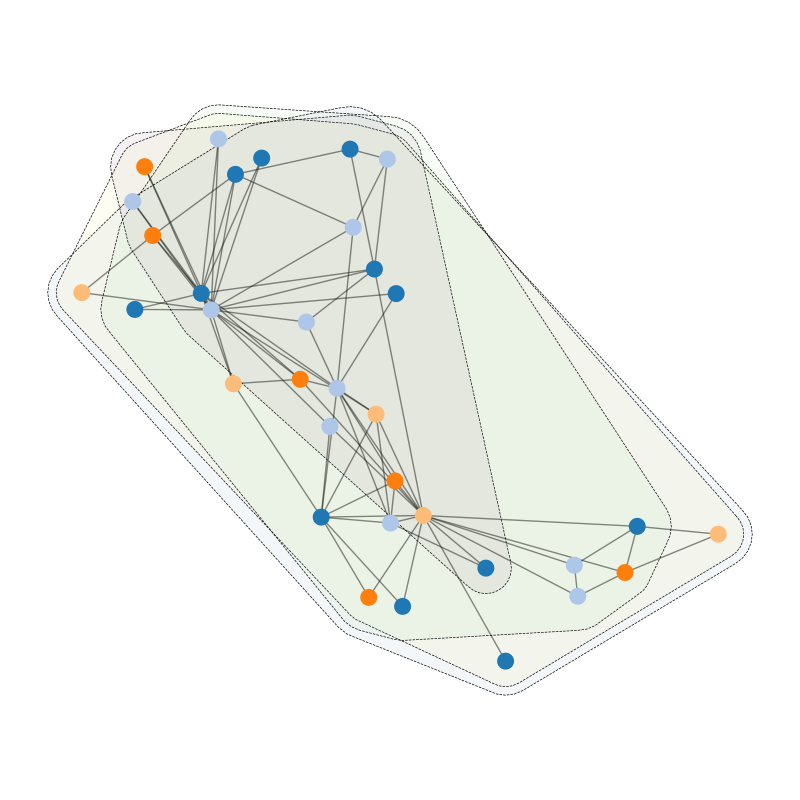}
}
\subfloat[louvain community]{
\label{fig:structure:b}
\includegraphics[width=0.24\textwidth]{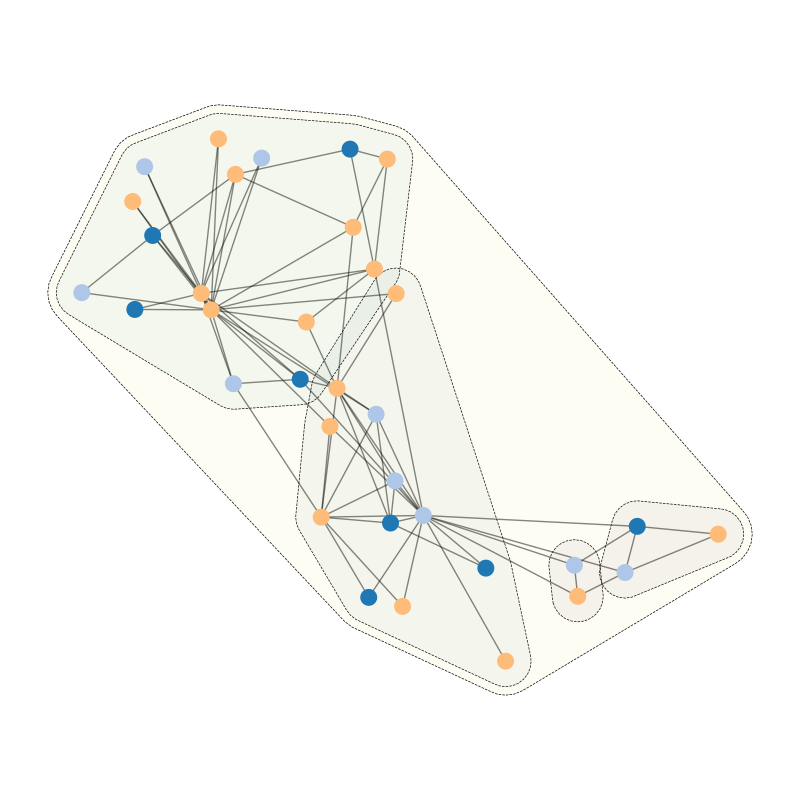}
}
\subfloat[one-hop neighbour]{
\label{fig:structure:c}
\includegraphics[width=0.24\textwidth]{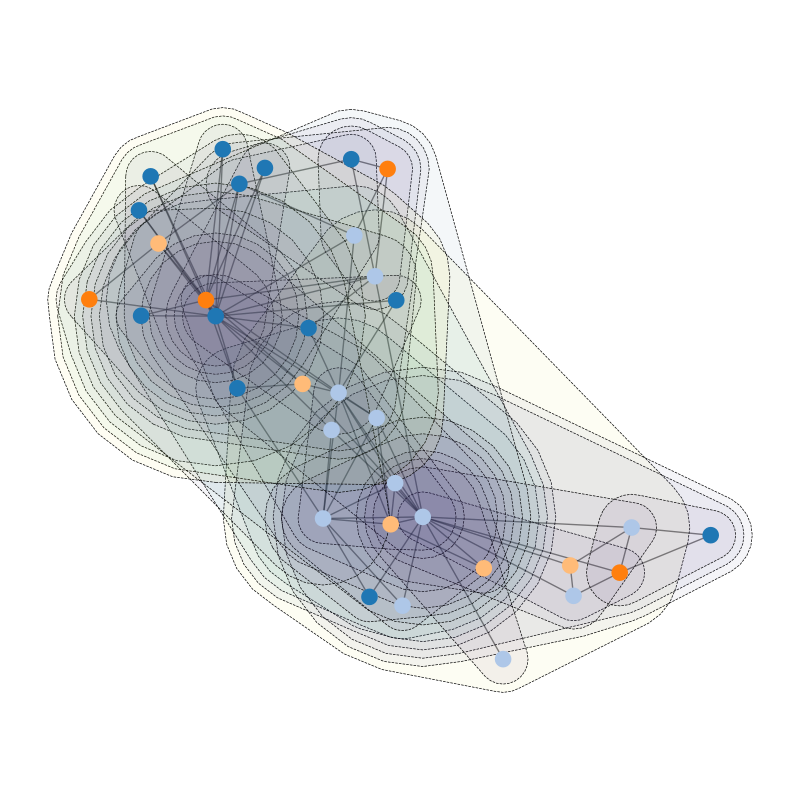}
}
\subfloat[Ours]{
\label{fig:structure:d}
\includegraphics[width=0.24\textwidth]{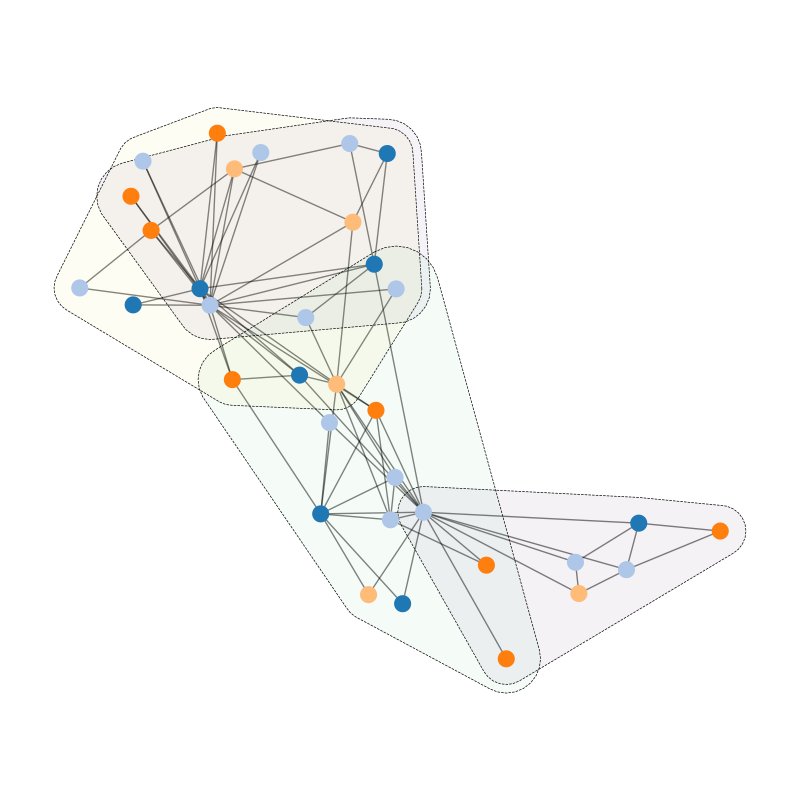}
}\vspace{-0.1in}
\caption{Hyperedge Structure }\vspace{-0.2in}
\label{fig:structure}
\end{figure*}

\vspace{-0.1in}
\subsection{Effectiveness on Social Environment Awareness}
\label{task:envaware}
To evaluate the hyperedge quality of our model, we present the hyperedge structure on the KaraClub dataset using our model, and the other three mostly used hyperedge construction methods including attribute clustering, community finding, and one-hop neighbor. Results are shown in Figure \ref{fig:structure}, from which we can find that the clustering-based method is usually not flexible to find diversified hyperedges. As presented in Figure \ref{fig:structure:a}, many hyperedges via clustering are heavily overlapped and this method also ignore the connectivity of pair-wise relations. Compared with the clustering-based method, the community-based method (Figure \ref{fig:structure:b}) can construct more hierarchical hyperedges because we 
can recursively call back the community finding model within a relatively large community to generate more sub-communities. However, this method often leaves many isolated nodes or generates some very small-sized groups, which may limit the hypergraph learning performance. Besides, hyperedge overlapping of this method is usually insufficient to support information diffusion on the hypergraph. According to Figure \ref{fig:structure:b}, neighbor-based hyperedges usually have much hyperedge overlapping and also keep relatively hierarchical structures. However, this method usually generates superfluous hyperedges, which may decrease efficiency. Compared with the above methods, our model integrates node attributes and their connectivity and the hyperedges are more concise than the neighbor-based method but still keep the better overlapping quality and hyperedge sizes than cluster-based and community-based methods.

\begin{figure}[h]
\centering
\includegraphics[width=0.28\textwidth]{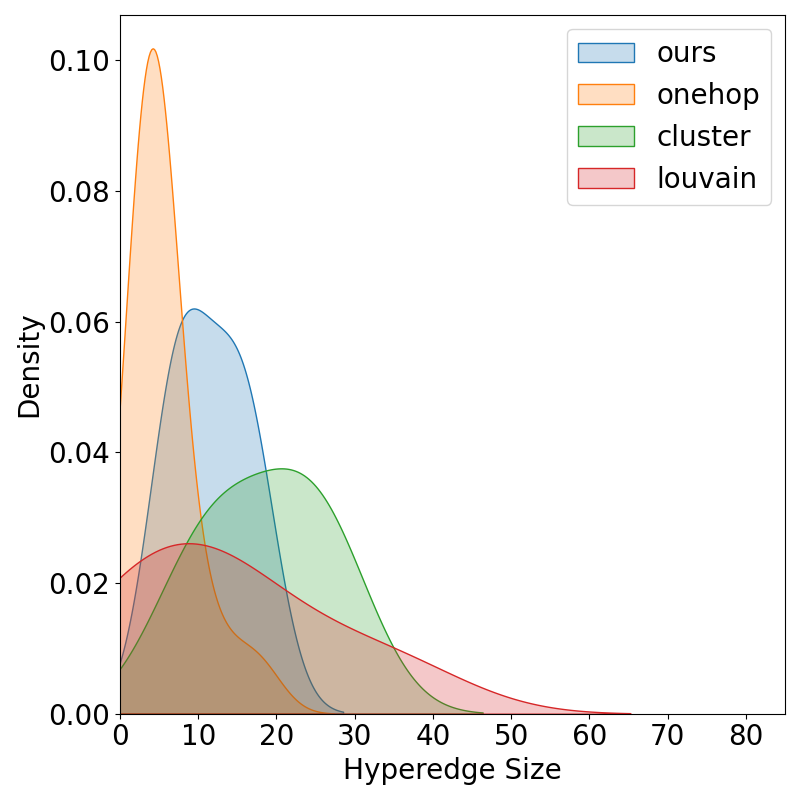}\vspace{-0.1in}
\caption{Hyperdge Distributions}\vspace{-0.1in}
\label{fig:hyperedge_dis}
\end{figure}

We further analyze the hyperedge size distribution on a larger facebook dataset and compare different hypergraph constructions in Figure \ref{fig:hyperedge_dis}, from which we can find that our method keeps a balanced distribution between one-hop neighbors and attribute clustering, and in the meanwhile present more significant long-tailed quality than Louvain community finding.

\vspace{-0.1in}
\subsection{Ablation Study}
\label{task:ablation}

\subsubsection{Convergence Analysis on Negative Contrastive loss}\label{subsec:robust}
\begin{figure}[h]
\centering
\includegraphics[width=0.3\textwidth]{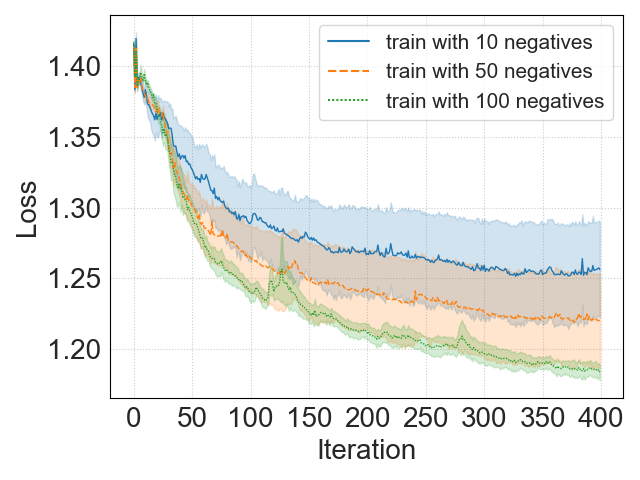}\vspace{-0.1in}
\caption{Training Loss w.r.t Negative Ratio}\vspace{-0.2in}
\label{fig:loss}
\end{figure}

We change negative ratios in our contrastive dual loss as one positive sample with 10, 50, and 100 negatives. Then we repeat each type training $10$ times and draw the loss reducing curve as shown in Figure \ref{fig:loss} where each dark colored line denotes the average loss value as training iteration increases and the shallow color band is the 65\% confidential band. From the figure, we can find that our model can efficiently come to convergence within 200 iterations with 10 negatives. When the negative ratio increased, the convergence state might come up a little late but can be still reached within 400 epochs. 

\subsubsection{Effectiveness Analysis on Main Components}
\begin{figure}[h]
\centering
\includegraphics[width=0.4\textwidth]{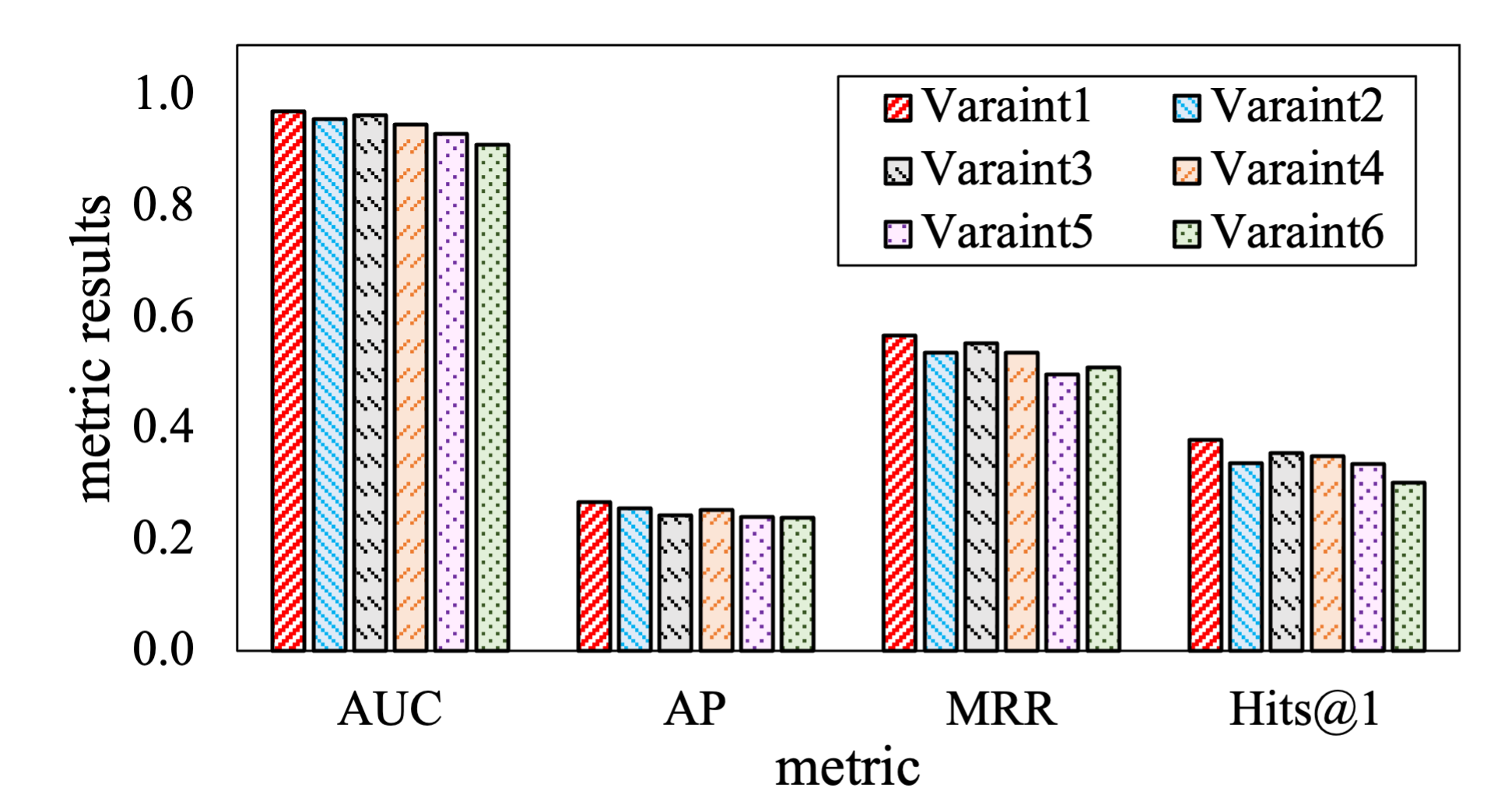}\vspace{-0.1in}
\caption{Effectiveness on Main Components}\vspace{-0.2in}
\label{fig:aba}
\end{figure}

To further demonstrate the effectiveness of our model, we remove several main components and compare the following model variants: Our full model with contrastive dual loss (Variant1) and normal reconstruction loss (Variant4); GCN backend alone with contrastive dual loss (Variant2) and normal reconstruction loss (Variant5); Hypergraph Convolution Network (HCN) backend alone with contrastive dual loss (Variant3) and normal reconstruction loss (Variant6). As shown in Figure \ref{fig:aba}, different backends with contrastive dual loss outperform the performance with normal reconstruction loss, which suggests the contribution of our proposed contrastive dual loss. We also find that our full model performs better than either the GCN backend or the HCN backend, which further demonstrates the effectiveness of our graph neural network model.

\vspace{-0.1in}
\subsection{Analysis on Social Equivalence and Conformity}
\label{task:equ_con}
\begin{figure}[h]
\centering
\subfloat[Social Equavelence.]{
\label{fig:equ_con:a}
\includegraphics[width=0.23\textwidth]{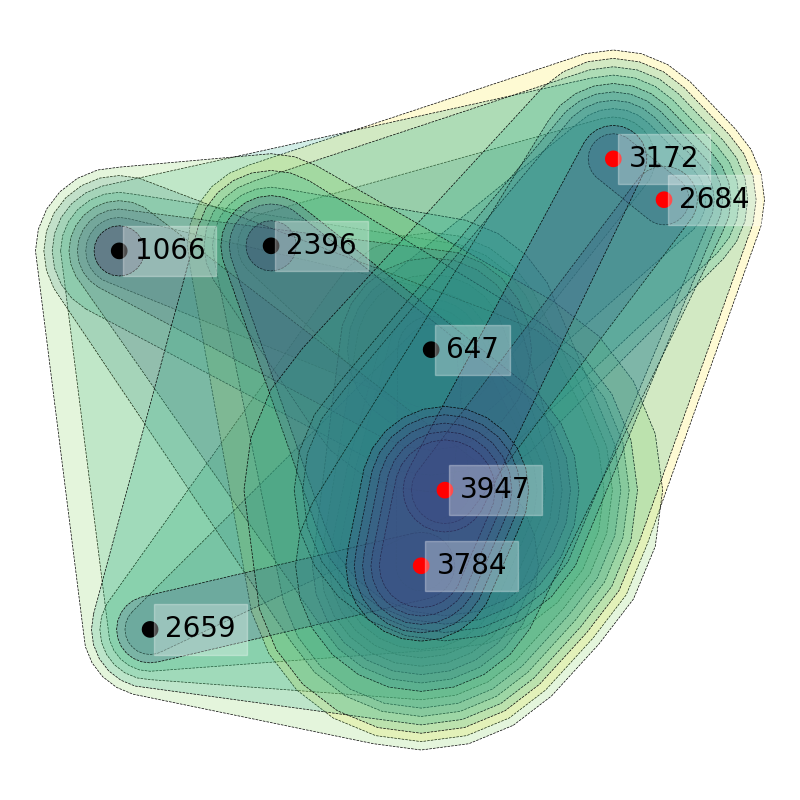}%
}
\subfloat[Social Conformity.]{
\label{fig:equ_con:b}
\includegraphics[width=0.23\textwidth]{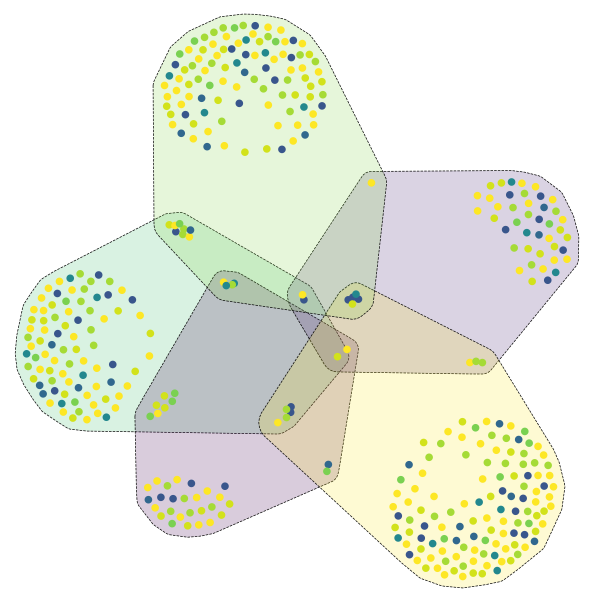}%
}\vspace{-0.1in}
\caption{Social Equivalence and Conformity by our Model}\vspace{-0.1in}
\label{fig:equ_con}
\end{figure}

In Figure \ref{fig:equ_con:a}, we visualize user environments in Facebook datasets where red nodes with closer distance indicate they are closely connected in testing data and black nodes with far distance are also far away in the real world. For a clear illustration, we did not show the other nodes. From this figure, we can find that those closely connected pairs are usually located in similar environments. On the contrary, users far away from each other usually have different environments. This observation indicates that our model can reflect the social equivalence and successfully integrate this sociological criterion into model learning.

\vspace{-0.1in}

In Figure \ref{fig:equ_con:b}, we present five groups (hyperedges) and their members on the PersonalityCafe dataset. Here yellow color denotes the majority of personality labels in a hyperedge, the green color is similar to personality labels, and other dark colors denote the distinct labels. From this figure we can observe that most members in a found hyperedge intend to be similar w.r.t their personality traits, suggesting the natural social conformity during the model learning.

\vspace{-0.1in}
\subsection{Simulated Social Environment Evolving}
\label{task:evol}
\begin{figure}[h]\vspace{-0.2in}
\centering
\includegraphics[width=0.33\textwidth]{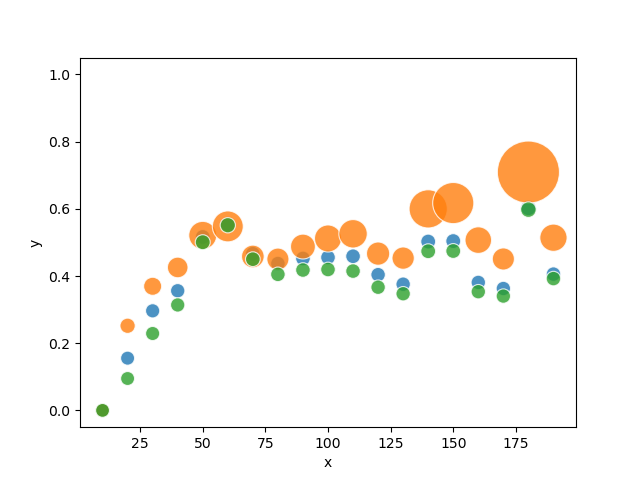}\vspace{-0.05in}
\caption{Simulated Analysis on Group Evolving. 
}\vspace{-0.1in}
\label{fig:evolving}
\end{figure}

We select 3 groups in the Facebook dataset and present the simulated group evolving patterns in Figure \ref{fig:evolving} where the x-axis denotes the evolving stage, which is simulated by different learning epochs; the y-axis denotes the ratio of significant node number out of the original nodes in the hyperedge. Here the ``significant node'' means we first use user representations at different stages to multiply the hyperedge representation (such as the average of the total members); then we get a probability of this user belonging to this group; at last we select the significant node when the probability is larger than $0.5$. The size of each bubble denotes the absolute number of significant nodes at a specific step. The orange hyperedge has very few connections to other hyperedges (less overlapping), while the green and the blue hyperedges have many connections to other hyperedges (more overlapping). 
For clear illustration, we only present the selected hyperedges, from which we can observe that the orange group increasingly attracts its members to be more loyal because it is more irreplaceable than the other groups. The rest groups have more overlapping with the other groups, making their positions more alternative. Therefore the ratio of ``loyal'' members keeps stable after several evolving stages.

\vspace{-0.1in}
\subsection{Analysis on Social Polarization}
\label{task:pola}

\begin{figure}[h]
\centering
\subfloat[Initial Stage]{
\label{fig:l1}
\includegraphics[width=0.22\textwidth]{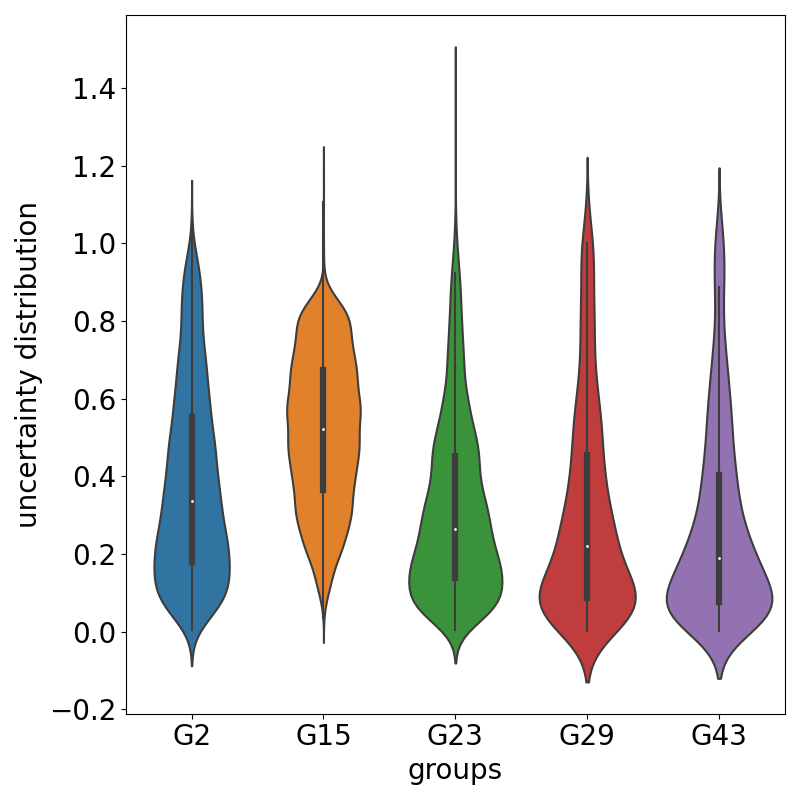}
}
\subfloat[Updated Stage]{
\label{fig:l3}
\includegraphics[width=0.22\textwidth]{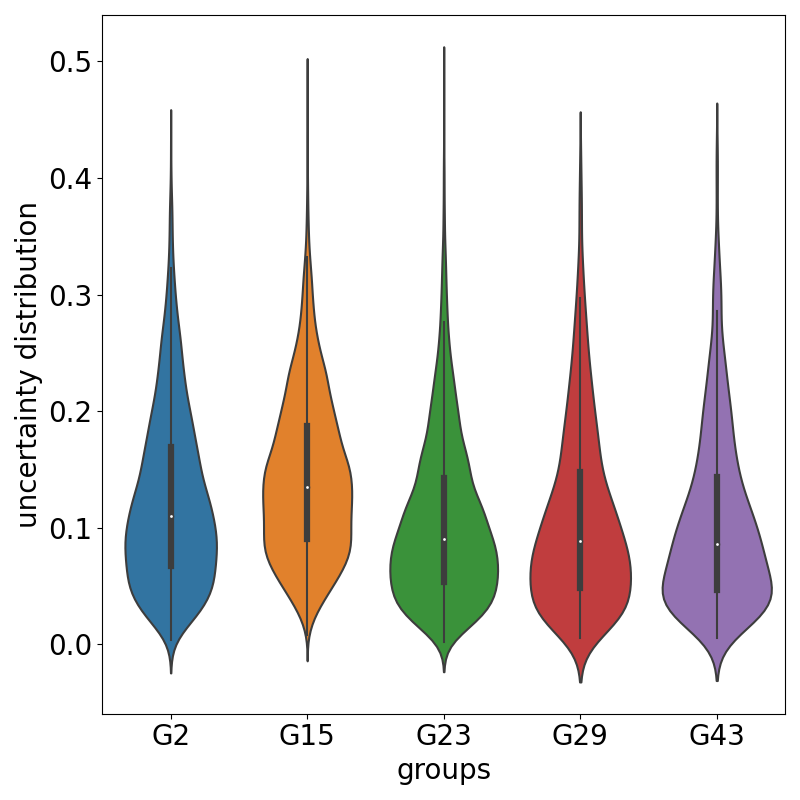}
}\vspace{-0.1in}
\caption{Social Polarization: An Entropy Perspective.}\vspace{-0.1in}
\label{fig:social_pola}
\end{figure}

Here we randomly selected five groups on the PersonalityCafe and then we draw the member uncertainty distribution within each group as shown in Figure \ref{fig:social_pola}. The overall uncertainty in a group can be measured by the group entropy shown in Table \ref{tab:entropy}. The higher group entropy means this group members are mostly not that confident about their belief in the group. On the contrary, the lower group entropy denotes the significance of social polarization. 

\vspace{-0.2in}

\begin{table}[h]
\centering
\caption{Comparison of Group Entropy}\vspace{-0.1in}
\label{tab:entropy}
\resizebox{0.25\textwidth}{!}{%
\begin{tabular}{@{}l|lllll@{}}
\toprule
        & G2       & G15       & G23      & G29      & G43      \\ \midrule
Initial & 963 & 2820 & 896 & 222 & 218 \\
Updated & 443 & 1166 & 404 & 110 & 113 \\ \bottomrule
\end{tabular}
}\vspace{-0.2in}
\end{table}

\section{related works}
Online behavior analysis plays a crucial role in many web-based applications such as recommendation \cite{xia2021self}, network alignment \cite{trung2020adaptive}, anomaly detection \cite{tam2019anomaly,tang2022rethinking}, etc. Since most online users interact with each other via various social relations, there are many works trying to organize them as graphs, leading to a huge number of graph-based models applied. In recent years, the mainstream models are graph neural networks such as GCN \cite{kipf2017semi}, GAT \cite{velivckovic2017graph}, GRN \cite{dai2018learning}, and their extensions for heterogeneous \cite{wang2019heterogeneous} and dynamic cases \cite{li2019predicting}. Usually, graph neural network aims at learning low-dimensional representations for graph nodes, which can support various downstream tasks. In particular, for some graph-level tasks \cite{dou2021user}, we usually take graph pooling operation \cite{mesquita2020rethinking} on nodes and get the graph-level representation; for edge-level tasks, we can design a transition \cite{qu2020continuous} on node pairs to get the edge embedding; for node-level tasks, the learned node representation can be directly sent to downstream component \cite{wang2019heterogeneous}. These task types cover the most common situations in online behavior analysis such as fake news detection \cite{dou2021user}, recommendation \cite{xie2021graph}, link prediction \cite{qu2020continuous}, node classification \cite{velivckovic2017graph} et al., making the graph structural data widely used in this field. Beyond normal pair-wise relations, online social networks also have rich higher level relations which might be not directly observed. Recently, hypergraphs \cite{li_hyperbolic_2021,zhang2020hypergraph} have attracted more attention because the hyperedge can connect multiple nodes, which is perfect to learn more latent higher-order interactions. Unfortunately, most social networks do not directly present their hypergraph structure, leaving the hyperedge awareness an open problem \cite{sun2021multi,an2021hypergraph}. Even with effective hypergraph models, we can only learn the interaction between users and hyperedges \cite{feng2019hypergraph} alone, which is insufficient to learn many complicated patterns in sociological criteria. Besides data mining techniques, there are lots of advanced studies in sociological areas \cite{Freese2017}, which can be very promising to further enhance the data mining research. However, sociological studies are very broad and it is very crucial to apply clearly defined theories and quantitative methods to understand the social world because online data analysis might be severely restricted if they are based on a flawed understanding of sociological theory. Some recent sociological works such as Markovsky and Webster \cite{Markovsky2015theorycon, webster2022unequals} further systematically study the proper sociological theory construction.

\vspace{-0.1in}

\vspace{-0.1in}
\section{Discussion and Conclusion}\label{sec:con}

In this paper, we propose an effective social environment awareness method that can go beyond classic pair-wise relations. The social environments combine pair-wise relations, hyperedges, and the line graph to the hypergraphs with a novel social influence flowing model that can preserve rich sociological criteria. We further propose a package solution to evaluate widely concerned sociological criteria. Extensive experiments from both the data mining field and the sociological field strongly support the effectiveness of our framework. It is worth noting that sociological research on user behavior is a wide research field and there are more thorough theories proposed recently. Data analysis might be limited if it is only based on some partial understanding of sociological theory. Through our effort in this paper, we wish to inspire more advanced work from other researchers that can fuse more thorough sociological theory with data analysis techniques.

\vspace{-0.1in}
\section*{Acknowledgments}
The work was supported by grants from the Research Grant Council of the Hong Kong Special Administrative Region, China (Project No. CUHK 14217622), NSFC Grant No. U1936205 and CUHK Direct Grant No. 4055159. National Key Research and Development Program of China (2022YFB4500300). NSFC Grant No. 61972087. Australian Research Council Future Fellowship (Grant No. FT210100624) and Discovery Project (Grant No. DP190101985). \dotuline{The first author, Dr. Xiangguo Sun, in particular, wants to thank his parents for their kind support during his tough period.}

\vspace{-0.2in}
\bibliographystyle{IEEEtranN}

\normalem 
\bibliography{ref}

\begin{IEEEbiography}[{\includegraphics[width=1in,height=1in,clip,keepaspectratio]{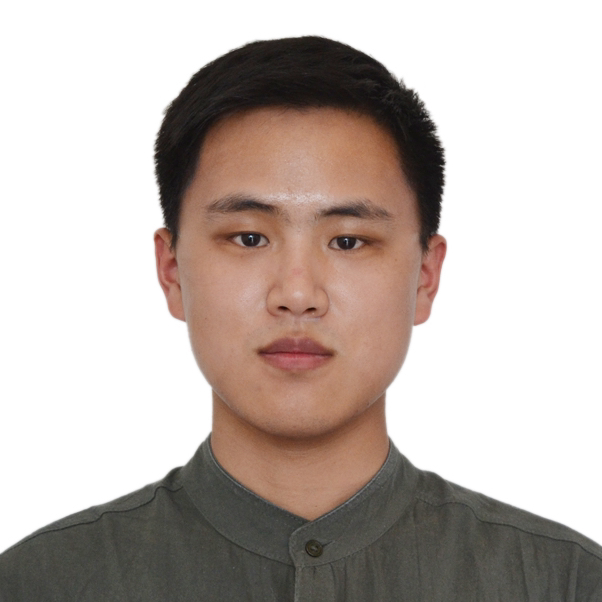}}]{Xiangguo Sun}
 works as a postdoctoral research fellow at the Chinese University of Hong Kong and works with Prof. Hong Cheng. He is also a visiting researcher at the Research Institute of Artificial Intelligence in Zhejiang Lab hosted by Dr. Hongyang Chen. He received his Ph.D. from Southeast University in Jan 2022 under the supervision of Prof. Bo Liu. During his Ph.D. study, he worked as a research intern in Microsoft Research Asia with Dr. Hang Dong and Bo Qiao, and visited the University of Queensland as a research scholar hosted by Prof. Hongzhi Yin. His research interests include social media analytics and hypergraph learning. His work has been published in some of the most prestigious venues such as SIGKDD, VLDB, TKDE, TNNLS, The Web Conference (WWW), TOIS, WSDM, CIKM etc.
\end{IEEEbiography}
\vspace{-10 mm}

\begin{IEEEbiography}[{\includegraphics[width=1in,height=1in,clip,keepaspectratio]{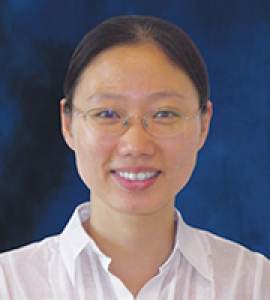}}]{Hong Cheng} is a Professor in the Department of Systems Engineering and Engineering Management, Chinese University of Hong Kong. She received the Ph.D. degree from the University of Illinois at Urbana-Champaign in 2008. Her research interests include data mining, database systems, and machine learning. She received research paper awards at ICDE'07, SIGKDD'06, and SIGKDD'05, and the certificate of recognition for the 2009 SIGKDD Doctoral Dissertation 
 Award. She received the 2010 Vice-Chancellor's Exemplary Teaching Award at the Chinese University of Hong Kong.
\end{IEEEbiography}

\vspace{-10 mm}

\begin{IEEEbiography}[{\includegraphics[width=1in,height=1in,clip,keepaspectratio]{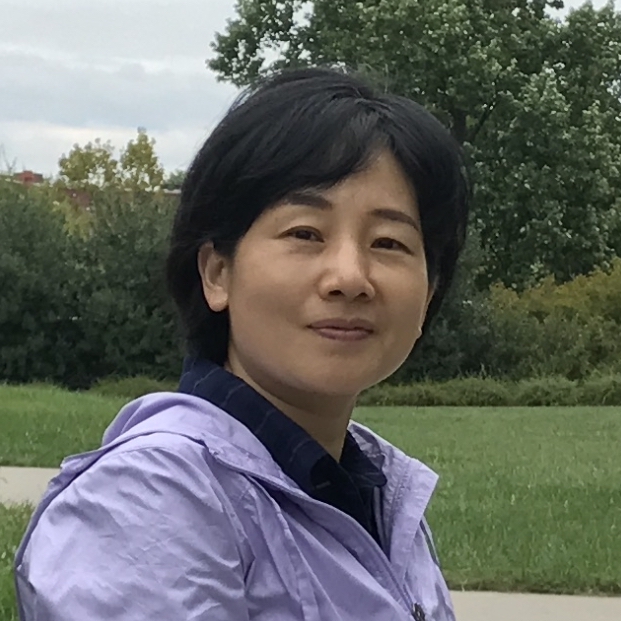}}]{Bo Liu} is a full professor at the School of Computer Science and Engineering, Southeast University, China. She received her doctoral degree from Southeast University under the supervision of Chief Scientist Prof. Junzhou Luo. She is the leader of three NSF projects of China and has published more than 60 papers in reputed journals and conferences including WWW, TKDE, WWWJ, TOIS, ToN et al. Her research interests include anomaly detection, social community, social influence, and recommendation.
\end{IEEEbiography}
\vspace{-10 mm}

\begin{IEEEbiography}[{\includegraphics[width=1in,height=1in,clip,keepaspectratio]{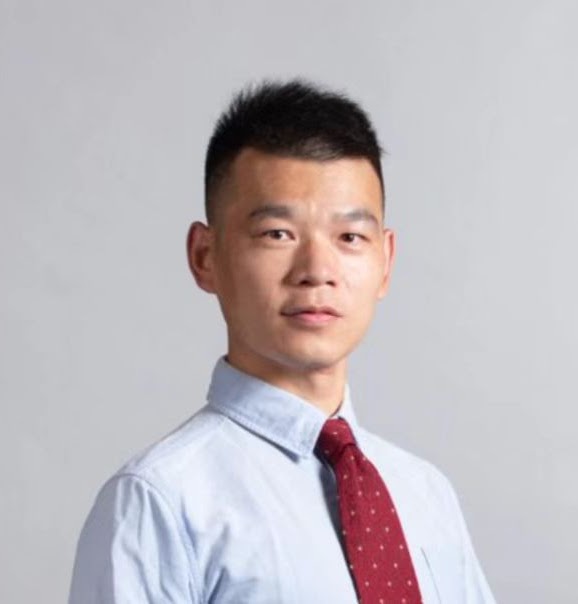}}]{Jia Li} is an assistant professor in HKUST (Guangzhou). He received Ph.D. degree at The Chinese University of Hong Kong in 2021. Before that, he worked as a full-time data mining engineer at Tencent from 2014 to 2017, and research intern at Google AI (Mountain View) in 2020. His research interests include graph learning and data mining. Some of his work has been published in TPAMI, ICML, NeurIPS, WWW, KDD and et al.
\end{IEEEbiography}
\vspace{-10 mm}

\begin{IEEEbiography}[{\includegraphics[width=1in,height=1in,clip,keepaspectratio]{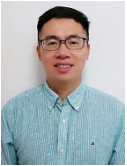}}]{Hongyang Chen} 
is the deputy director and senior research expert for the research center of graph computing, Zhejiang Lab, China. He was the Distinguished Lecturer of the IEEE Communication Society from 2021 to 2022. Prof. Chen is an adjunct professor at Hangzhou Institute for Advanced Study, The University of Chinese Academy of Sciences, and Zhejiang University, China. He received his Ph.D. degree from the University of Tokyo, Japan in 2011. He has published more than 120 top-tier papers related to data-driven intelligent systems, graph machine learning, big data mining, and intelligent computing. Most of them are published in SIGMOD, AAAI, TKDE, JSAC, CIKM, etc. He has more than 30 international patents, many of which have been adopted as international standards.
\end{IEEEbiography}
\vspace{-10 mm}

\begin{IEEEbiography}[{\includegraphics[width=1in,height=1in,clip,keepaspectratio]{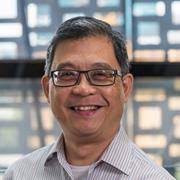}}]{Guandong Xu} is a full professor and leader in the Data Science and Machine Intelligence Lab at the University Technology of Sydney. He received Ph.D. degree from Victoria University, Australia. 
He was the winner of the 2020 Global Efma-Accenture Insurance Innovation Award in Workforce Transformation and the Digital Disruptors Winner for Skills Transformation of Small Work Teams, Australian Computer Society (2019). He has more than 220 papers published in top journals and conferences such as TOIS, TNNLS, KDD, AAAI, ICDM, IJCAI and et al.
\end{IEEEbiography}
\vspace{-10 mm}

\begin{IEEEbiography}[{\includegraphics[width=1in,height=1in,clip,keepaspectratio]{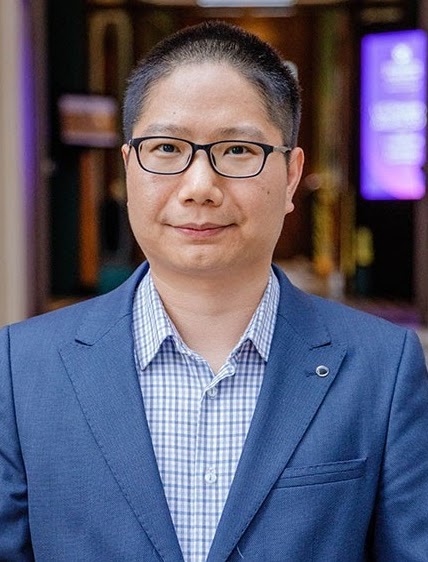}}]{Hongzhi Yin} works as ARC Future Fellow and associate professor with The University of Queensland, Australia. He was recognized as Field Leader of Data Mining \& Analysis in The Australian's Research 2020 magazine, the recipient of the 2022 AI 2000 Most Influential Scholar Honorable Mention in Data Mining (10 Years Impact), and featured among the World's Top 2\% Scientists List published by Stanford University (Single Year Impact) and AD Scientific Index 2022 (5 Years Impact). His current research interests include recommender systems, graph embedding, and mining, chatbots, social media analytics, and mining, edge machine learning, trustworthy machine learning, decentralized and federated learning, and smart healthcare. He has published 220+ papers with H-index 52, including 22 most highly cited publications in Top 1\% (CNCI), 120 CCF A and 70+ CCF B, 120 CORE A* and 70+ CORE A, such as KDD, SIGIR, WWW, WSDM, SIGMOD, VLDB, ICDE, AAAI, IJCAI, ACM Multimedia, ECCV, IEEE TKDE and TNNL, VLDB Journal and ACM TOIS. He has won 6 Best Paper Awards, such as Best Paper Award at ICDE 2019, Best Student Paper Award at DASFAA 2020, ACM Computing Reviews' 21 Annual Best of Computing Notable Books and Articles, and Best Paper Award Nomination at ICDM 2018 (being invited to be published in KAIS special issue of best papers).
\end{IEEEbiography}

\vfill
\end{document}